\documentclass[aps,pre,twocolumn,groupedaddress,showpacs,floatfix]{revtex4}
\usepackage{graphicx}
\begin{document}
\title{Master equation approach to friction at the mesoscale}

\author{O.M. Braun}
\email[E-mail: ]{obraun.gm@gmail.com}
\homepage[Web: ]{http://www.iop.kiev.ua/~obraun}
\affiliation{Institute of Physics, National Academy of Sciences of Ukraine,
  46 Science Avenue, 03028 Kiev, Ukraine}
\author{M. Peyrard}
\email[E-mail: ]{Michel.Peyrard@ens-lyon.fr}
\affiliation{Laboratoire de Physique de l'Ecole Normale Sup\'{e}rieure de Lyon,
  46 All\'{e}e d'Italie, 69364 Lyon C\'{e}dex 07, France}
\date{\today}
\begin{abstract}
At the mesoscale friction occurs through the breaking and formation of
local contacts. This is often described by the earthquake-like model
which requires numerical studies. We show that this phenomenon can
also be described by a master equation, which can be solved
analytically in some cases and provides an efficient numerical
solution for more general cases. We examine the effect of temperature
and aging of the contacts and discuss the statistical properties of
the contacts for different situations of friction and their
implications, particularly regarding the existence of stick-slip.
\end{abstract}
% 46.55.+d (Tribology and mechanical contacts)
% 81.40.Pq (Friction, lubrication and wear in materials science)
% 61.72.Hh (indirect dislocations,slip,creep,friction)
\maketitle

%=========================================================================
\section{Introduction}
\label{intro}

In spite of its crucial practical importance, friction is still far
from being fully explained \cite{P0,MUR2003,BN2006}. Besides a proper
explanation
of the static friction laws, the dynamical aspects of friction are even
less understood. This is exemplified by the
problem posed by the lack of a well established mechanism for
the familiar stick-slip phenomenon that one perceives with a door's
creak or the playing of a violin with a bow.
Many phenomena in nature, where one part of a system moves in contact with
another part, exhibit such a stick-slip motion which changes to smooth
sliding with the increase of the driving velocity
\cite{P0,MUR2003,BN2006,BC2006}.
Microscopically this phenomenon was first explained by Robbins and
Thompson \cite{RT1991} by a melting-freezing mechanism: a thin
lubricant film between the moving surfaces melts during slips and
solidifies again at sticks.  Such a behavior is typical for
conventional, or ``soft'' lubricants, when the lubricant-surface
interaction is stronger than the interaction between the lubricant
molecules \cite{BN2006}.  A ``hard'' lubricant, which remains in a
solid state during slips, also demonstrates the stick-slip to smooth
sliding transition which now emerges due to inertia
\cite{BN2006,BP2001}.  In both cases, however, the transition from
stick-slip to smooth sliding in a planar tribological contact
is found to occur
at a driving velocity $v_c \sim 10^{-2}c \sim 1-10$~m/s (where $c$ is
the sound velocity), which is more than six orders of magnitude higher
than experimentally observed \cite{BN2006,BR2002,BPBFV2005}.
This leads to the conclusion that microscopic mechanisms of stick-slip
have little relevance at the macroscopic level.

At the macroscopic scale, the stick-slip and smooth-sliding regimes
can be explained with a phenomenological theory
\cite{P0,BC2006,pheno,P1997}.  It is based on a ``contact-age
function'' which depends on the previous history of the system.  This
theory leads to an excellent agreement with experiments if the model
parameters are suitably chosen.  Unfortunately, this approach remains
purely phenomenological.  The corresponding equations cannot be
derived from a microscopic-scale analysis.

Another explanation is based on earthquake-like models (the EQ model)
such as the Burridge-Knopoff spring-block model \cite{BK1967,CL1989}
later developed by Olami, Feder, and Christensen \cite{OFC1992}.
It was adjusted to describe friction by Persson~\cite{P1995}
and then explored in a number of works~%
\cite{BR2002,FKU2004,FDW2005,BP2008,BT2009,BBU2009}.
In this model the two solid blocks touch one another at point
junctions, corresponding to asperities of the surfaces
which pin the relative position of the blocks (Fig.~\ref{A01}).
A local force, exerted by the moving top block, is associated to each
contact, as well as a threshold for breaking. The model describes
friction as resulting from the combined effects of the breaking of
contacts and their pinning again after the release of their internal
stress.
Computer simulations of a simplified variant of the EQ model,
where all contacts have the same threshold
\cite{P1995,BR2002},
showed that
the earthquake-like model reproduces the experimentally observed transition
from stick-slip to smooth sliding,
provided the following assumptions are made:
(\textit{i}) the model must be two-dimensional,
(\textit{ii}) the spatial distribution of contacts  must be random,
(\textit{iii}) it should exist an interaction between the contacts,
and
(\textit{iv}) the model must incorporate an increase of the static threshold
 on the time of stationary contact similarly to
the ``age-function'' assumption of the phenomenological theory. % approach.
Recently~\cite{FDW2005,BP2008} it became clear that a crucial ingredient
of the EQ-like model in tribological applications is the
distribution of static yield thresholds for the breaking of
individual contacts. In the early variants of the model, all
contacts had been assumed to be identical for the sake
of simplicity, and a distribution of thresholds appeared
only implicitly due to temperature fluctuations~\cite{P1995}
or due to interaction between the contacts~\cite{BR2002}. As a
matter of fact, the EQ model with identical contacts is
a singular case~\cite{FDW2005}. It admits a periodic solution which
can be interpreted as a form of stick-slip. Nonetheless
this solution remains largely unphysical since for example
it ceases to exist as soon as nonequivalent contacts are
considered, whatever their precise properties. As soon as
a finite width distribution of yield thresholds is taken
into account, the solution of the model in the quasi-static
limit always approaches a physical solution with smooth
sliding~\cite{FDW2005,BP2008}.
Incorporating a threshold
distribution  that does not evolve by the breaking and re-formation of
the contacts allows one to find the steady-state solution of
the EQ model analytically, and more importantly to
find conditions for appearance of the elastic instability,
which is the necessary condition for the stick-slip to
emerge~\cite{BP2008,BC2006}.

The approach based on the earthquake-like model has, however, two weak points.
First, most results can only be obtained with  computer simulation.
Second, the nature of contacts is not clearly specified.
While for two rough solid surfaces
the contacts may be associated with real asperities,
for two ideal flat mica surfaces in the
surface force apparatus (SFA) experiments \cite{YCI1993}
the nature of contacts is not clear.
Persson \cite{P1995} supposed that they may correspond to ``solid
islands'' in the fluidized lubricant,
but such an assumption remains on a speculative level.

Therefore, whether one looks at friction from the macroscopic or from
the microscopic viewpoint, the theoretical understanding is not
satisfactory. What is lacking is a theory that would not separately
consider the
two extreme limits, macroscopic or microscopic, but couple
them in a mesoscale approach. Our goal in the present work is to
explore an intermediate approach, which starts from the properties of
individual contacts and deduces macroscopic laws from an analytical
description based on a statistical analysis. It is interesting to
notice that the introduction of statistics in the theory of friction
\cite{GW1966} already allowed a significant progress in the
understanding of {\em static} friction, but here we are concerned with
{\em dynamical aspects} which enter through the continuous breaking and
re-forming of many local contacts. We introduce a master equation (ME)
which describes this phenomenon and couples local events with
macroscopic properties. It can be solved analytically in cases which
are particularly relevant, or studied numerically in more complex
situations, much more efficiently than with simulations of discrete EQ
models. Its major interest is to split the analysis in two independent
parts: (\textit{i}) the calculation of the friction force given by the master
equation provided the statistical properties of the contacts are
known, and (\textit{ii}) the study of the statistical properties of
the contacts, which may need inputs from the microscopic scale.

A preliminary report of the results have been presented in a short
letter \cite{BP2008}. In this paper we discuss
the master equation approach more thoroughly and
explore some of the consequences of this new viewpoint on friction by
examining several issues such as temperature effects
(Sec.~\ref{temperature}) and the aging of the contacts
(Sec.~\ref{aging}). We discuss the origin of the statistical
properties of the contacts (Sec.~\ref{disc}). The paper begins by a
brief review of some results of the EQ model (Sec.~\ref{earthquakes})
and the presentation of the master equation approach and some of its
analytical solutions in Sec.~\ref{analytics}.

%==================================================================
\section{Earthquake-like model}
\label{earthquakes}

\begin{figure}[h] %[t] \bigskip
\includegraphics[clip,width=6cm]{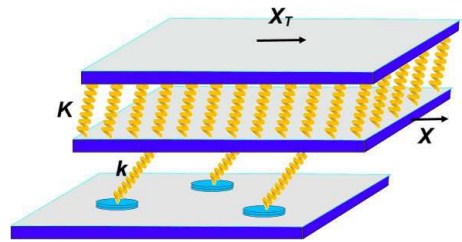} % clip % 8
\caption{\label{A01} (Color online):
The earthquake-like model. Friction occurs between the lower plate and
the top block schematized by two plates connected by a network of
springs to describe its internal shear stress. Thr contacts are
represented by the disks on the lower plate and their elasticity is
modeled by the springs connecting the lower plate and the bottom of
the top block. The arrow indicate the displacement of the block. $X$
and $X_T$ are the coordinates of the bottom and top of the block.
%[use A01]
}
\end{figure}

Let us begin with the earthquake-like model,
which belongs to the class of cellular automaton models.
This section, which follows earlier works \cite{P1995},
provides a reference for the master equation approach.
We use a variant of the Burridge-Knopoff spring-block model of
earthquakes adapted to tribology problems by Persson~\cite{P1995}.
The contacts form an array and
a local force $f_i (t)$ is associated with each contact.
All contacts are connected through springs of strength $k_i$,
corresponding to their shear elastic constant,
with the top block moving with a velocity $v$
and coupled frictionally with the fixed bottom block as shown in Fig.~\ref{A01}.
The contacts may also interact elastically between themselves.
As the top block moves, the surface stress at any contact increases,
$f_i (t) = k_i x_i (t)$,
where $x_i (t)$ is the shift of the $i$th junction from its
non-stressed position.
A single contact is assumed to be pinned whilst $f_i (t)<f_{si}$,
where $f_{si}$ is the static friction threshold for the given contact.
When the force % on a given contact
reaches $f_{si}$, a rapid local slip takes place,
during which the local stress in the block drops to the value $f_{bi}
\approx 0$ while
the elastic energy stored in the junction is released in the form of
phonons into the bulk.
Then the contact is pinned again, and the whole process repeats itself.
Let $N_c$ be the number of contacts (asperities, bridges, etc.) at the
interface.
Each contact is characterized by its area $A_i$.
The value of $k_i$, according to Persson \cite{P1995}, can be estimated as
$k_i \sim \rho \, c^2 \sqrt{A_i}$, where $\rho$ is the mass density and
$c$ is the transverse sound velocity of the material which forms the contacts
(see Appendix~\ref{shearkc}).
Let us assume that the substrates are rigid, i.e.,
the elastic constant of the substrate is infinite, $K=\infty$,
and also neglect elastic interactions between the contacts
through the substrate %, $k_{ij}=0$
as it was done in the statistical
studies of static friction \cite{GW1966}.

Let $P_c (x_s)$ be the normalized probability distribution
of values of the thresholds
$x_{si}=f_{si}/k_i$ at which % ``newborn'' (``fresh'')
contacts break, where $f_{si}$ is the maximum force that contact $i$
can sustain.
If we denote by $ \bar{x}_s$ its average value and by $\sigma_s$ its
standard deviation,
a typical example is the Gaussian $P_c (x) = G(x; \bar{x}_s,\sigma_s)$, where
\begin{equation}
G (x; \bar{x},\sigma)) = \frac{1}{\sigma \sqrt{2\pi}} \,
\exp \left[ -\frac{1}{2} \left( \frac{x - \bar{x}}{\sigma} \right)^2 \right].
\label{Q14e}
\end{equation}
In numerics, the distribution $P_c (x)$ is defined on the interval
$0 < x < x_m$, where $x_m \gg \bar{x}_s$
so that we use the corrected distribution
$P_c (x) = {\cal N} \left\{ P_G (x) - P_G (0) - (x/x_m) \left[ P_G
  (x_m) - P_G (0) \right] \right\}$,
where ${\cal N}$ is the normalization constant,
so that $P_c (x)$ satisfies the condition $P_c (0)=P_c (x_m)=0$.

To describe the kinetics of the model,
we introduce the distribution $Q(x;X)$ of the stretchings $x_i$
when the top substrate is at a position $X$. It is normalized by
\begin{equation}
\int_{-\infty}^{\infty} dx \, Q(x;X) = 1.
\label{Q5c}
\end{equation}

Let all contacts start from an initial distribution $Q(x;0)=Q_{\rm
  ini} (x)$ corresponding to the Gaussian one, $Q_{\rm ini} (x) = G(x;
\bar{x}_{\rm ini},\sigma_{\rm ini})$ with $\bar{x}_{\rm ini} \ll
\bar{x}_s$ and $\sigma_{\rm ini} \alt \sigma_s$.
Let us now apply an adiabatically increasing force $F$ to the top substrate,
while the bottom substrate remains fixed.
The force will induce a displacement $X$ of the top substrate.
According to third Newton's law (the law of action and reaction), in the
adiabatic regime of quasi-equilibrium $F$
must be compensated by the sum
of elastic forces in the contacts,
\begin{equation}
F(X)=N_c \langle k_i \rangle \int_{-\infty}^{\infty} dx \, x \, Q(x;X) \,,
\label{Q6}
\end{equation}
where we assumed that
$k_i$ and $x_i$ are independent random variables.

With the increase of $F$ and consequently $X$,
the stretching of a contact $i$ grows by the value $X$
with respect to its initial value,
until it reaches the threshold stretching $x_{si}$ for the given contact.
At this point the contact slides and $x_i$ drops;
we assume that the sliding is rapid and that
the stretching drops to zero, $x_{bi} = 0$.
In the numerical algorithm,
the contact closest to the threshold is found and the displacement necessary
to provoke a slip in this contact is added to all contacts.
When the contact $i$ reaches the threshold and $x_i$ is set to zero,
a new threshold is assigned to this contact randomly
from the distribution $P_c (x)$.
Then the whole process repeats itself.

\begin{figure}[h] %[t] \bigskip
\includegraphics[width=8cm]{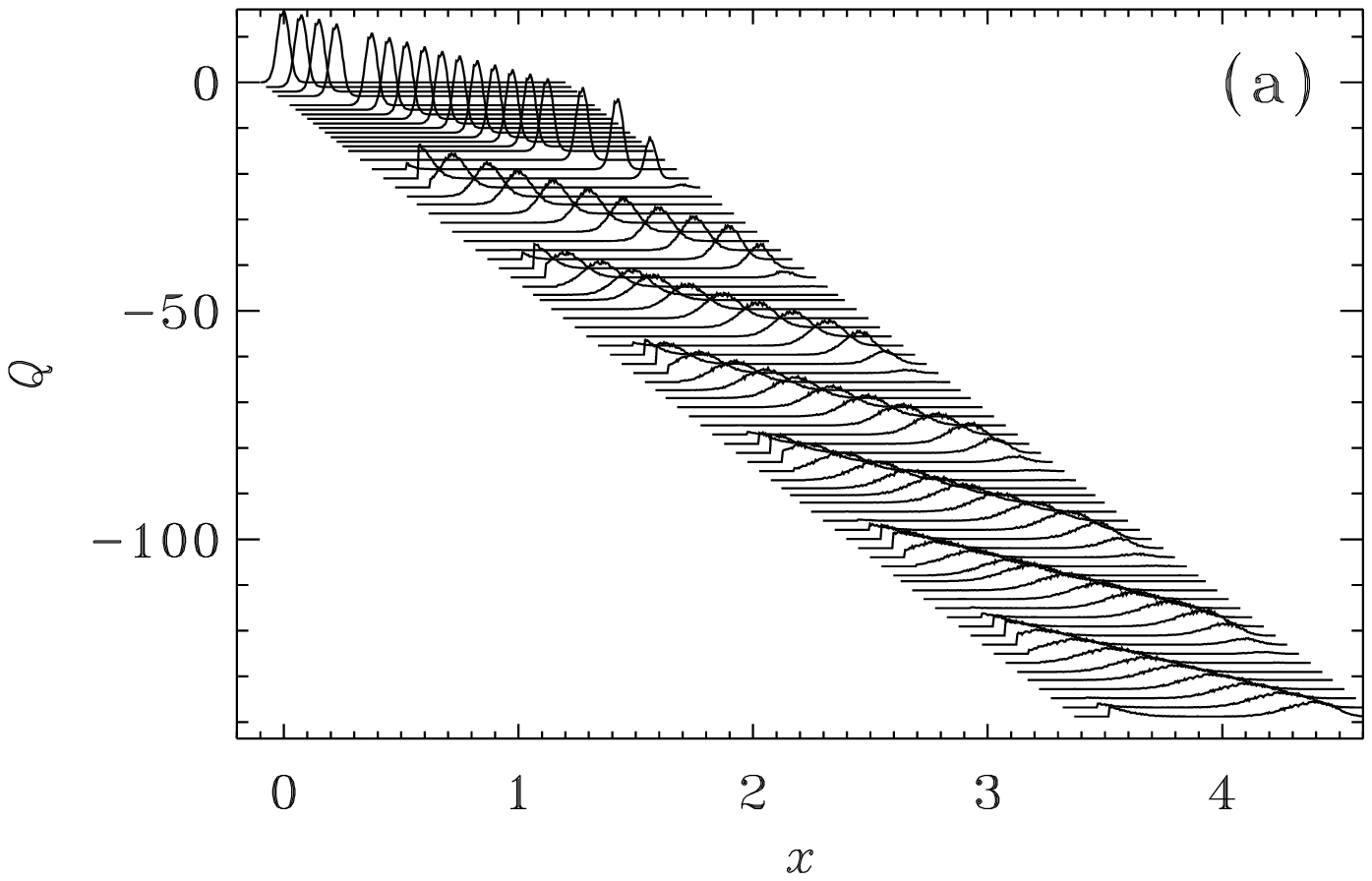} % clip
\includegraphics[width=7.8cm]{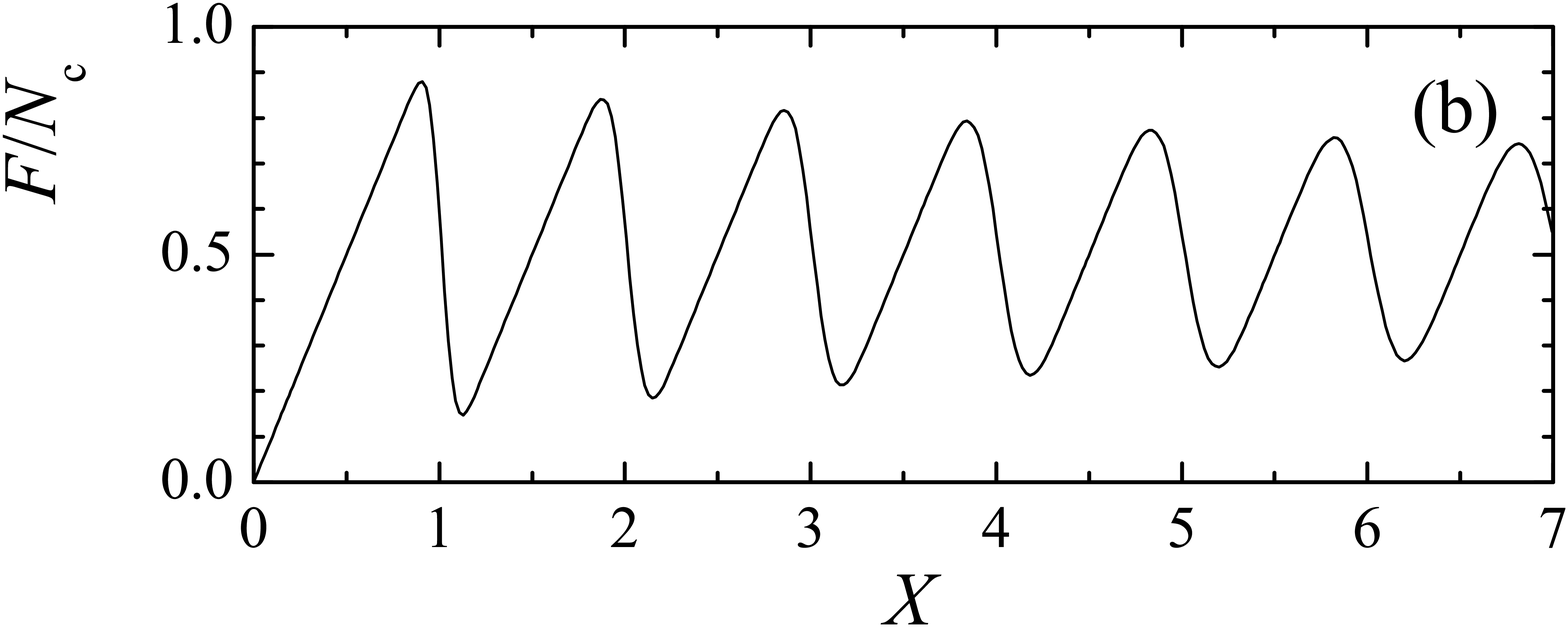}
\caption{\label{A02a}
(a) Short-time evolution of the EQ model.
Solid curves show the distribution $Q(x;X)$ for incrementally
increasing values of $X$
(with the step $\Delta X = 0.05$).
The distribution $P_c (x)$ is Gaussian with
$\bar{x}_s =1$ and $\sigma_s =0.05$,
the initial distribution $Q_{\rm ini} (x)$ is
Gaussian with $\bar{x}_{\rm ini} =0$ and $\sigma_{\rm ini} =0.025$.
For clarity each graph of  $Q(x;X)$ is shifted by $0.025$ along the
$x$ direction and by $-1$ along the $Q$ direction with respect to
the previous one.
(b) The corresponding dependence $F(X)$.
The elastic constants are $\langle k_i \rangle =1$ and the top block
is assumed to be fully rigid.
% [use dir A02new-short = A02a + evo-short.eps]}
}
\end{figure}
\begin{figure}[h] %[t] \bigskip
\includegraphics[width=8cm]{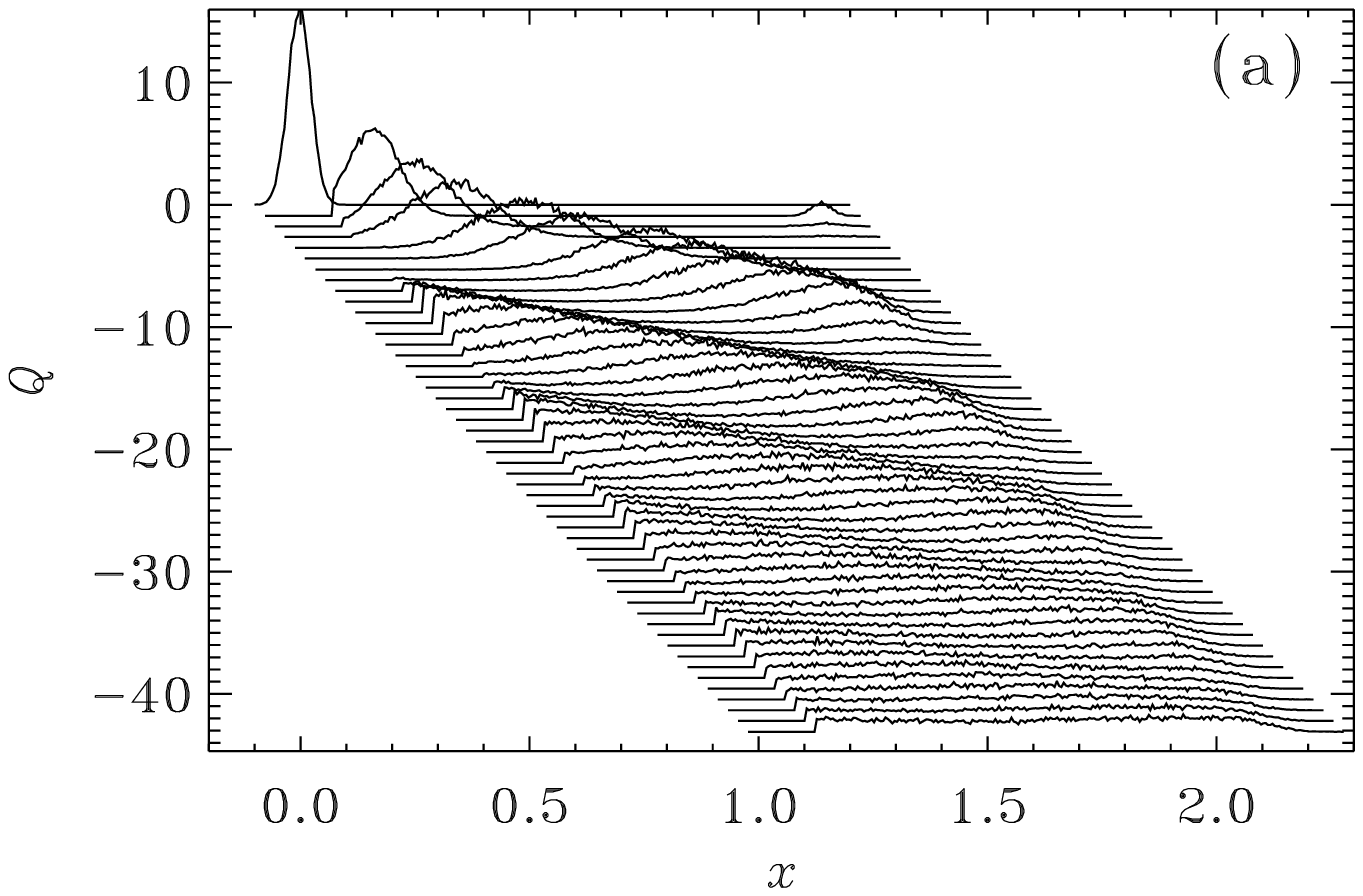} % clip
\includegraphics[width=8.2cm]{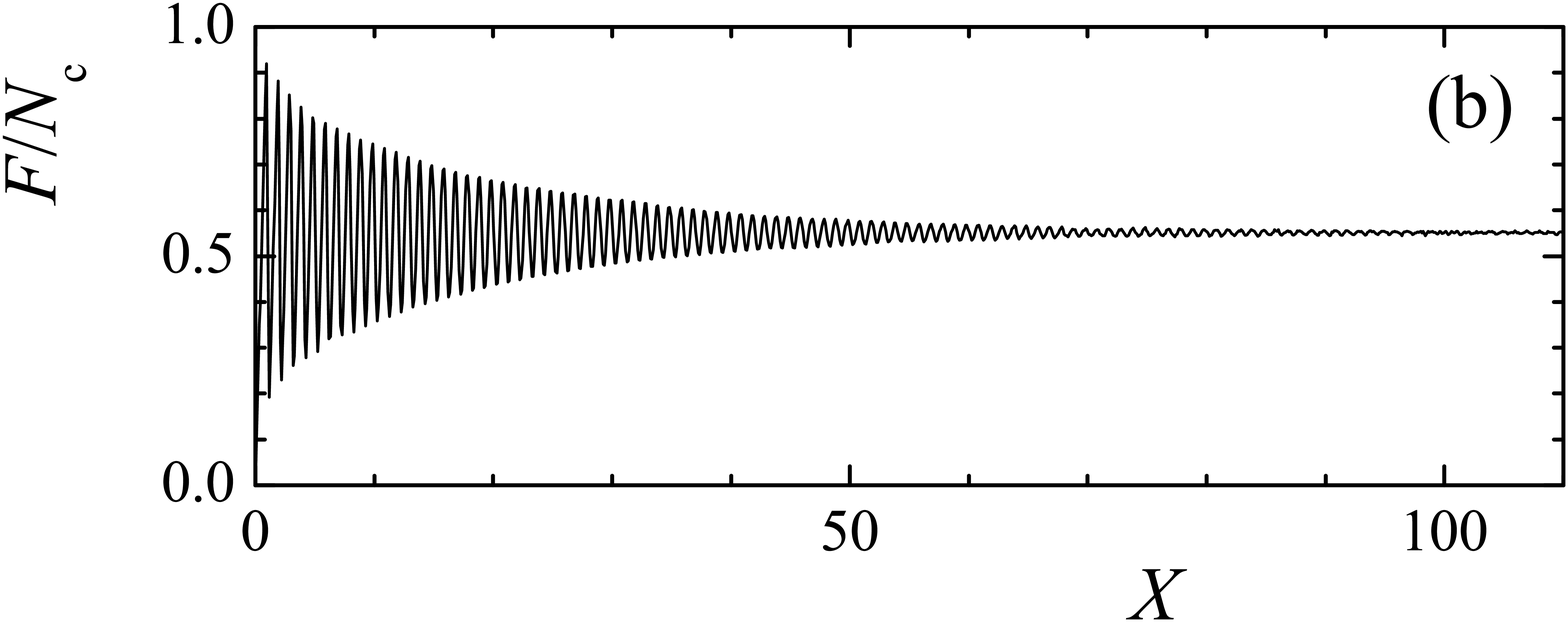}
\caption{\label{A02b}
The same as in Fig.~\ref{A02a} for long times
with the increment $\Delta X \approx 1.05$.
While the initial distribution extended to negative
    $x$ values, according to the model this is not the case for
    further ones, which gives rise to what appears as a bump at $x$ =
    0 (taking into account the shifts introduced for clarity).
%[use MP PRL = A02b + evo-long.eps]
}
\end{figure}
\begin{figure}[h] %[t] \bigskip
\includegraphics[clip, width=8cm]{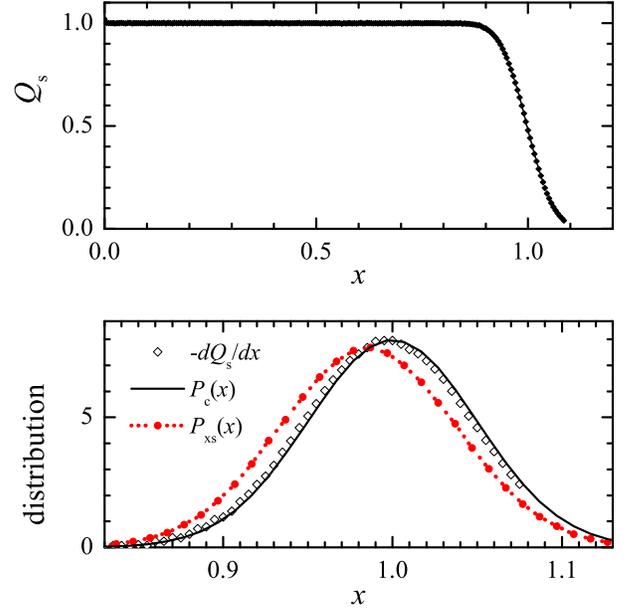}
\caption{\label{A02c}(Color online):
The averaged final distribution $Q_s (x)$ for the parameters used in
Figs.~\ref{A02a} and~\ref{A02b}.
The lower panel compares the functions $P_c (x)$ (solid line) and
$-dQ_s (x)/dx$ (open diamonds).
Red dotted curve and circles show the averaged final distribution of
static thresholds.
%[use A02c/FPQ-ori8.opj]
}
\end{figure}

The evolution of the system is shown in Fig.~\ref{A02a} for short times
and in Fig.~\ref{A02b} for long times.
One can see that, as time goes,
the initial distribution approaches a stationary distribution $Q_s (x)$,
where the total force $F$ becomes independent on $X$, and
the final distribution (see Fig.~\ref{A02c}) is independent of the initial one.
An elegant mathematical proof of this statement for a simplified
version of the EQ model was presented in Ref.~\cite{FDW2005};
the statement is valid for any distribution $P_c (x)$
except the singular case of $P_c (x) = \delta (x-x_s)$.

Note that the probability distribution $P_c (x)$,
which determines the value of the static threshold $x_{si}$ for
newborn contacts
is different from the concrete realization of the distribution of
static thresholds
$P_{xs} (x)$ (i.e., the histogram calculated over the array $\{x_{si}\}$).
While at the beginning $P_{xs} (x) = P_c (x)$,
then the function $P_{xs} (x)$ changes with time and finally takes the form
shown in Fig.~\ref{A02c} (lower panel) by solid curve and circles.
This is clearly demonstrated in Fig.~\ref{A02d} of Appendix~\ref{simple},
where we consider a simple case
of rectangularly-shaped function $P_c (x)$.

%------------------------------------------------------------------
\section{Master equation approach}
\label{analytics}

\subsection{Master equation}

Let us now introduce an analytical description of the evolution of the
distribution $Q(x;X)$.
When the top substrate is displaced by a small amount $\Delta \! X >0$
(the case of $\Delta \! X <0$ is considered in Appendix~\ref{loop}),
this increases all displacements of the contacts by $\Delta \! X$ too.
This displacement leads to three kinds of changes in the
distribution $Q(x;X)$: first, there is a global shift for all contacts,
second, some contacts break because their stretchings become too large,
and third, those broken contacts will form again,
at a lower stretching after a slip at the scale of those contacts,
which locally reduces the tension within the top substrate
(or the tension of the corresponding springs in the spring-block model).
These three contributions can be written as a master equation for $Q(x;X)$:
\begin{equation}
Q(x;X + \Delta \! X) = Q(x - \Delta \! X;X) - \Delta Q_-(x;X) + \Delta Q_+(x;X).
\label{Q2}
\end{equation}

The first term in the r.h.s.\ of Eq.~(\ref{Q2}) is just the shift.

The second term $\Delta Q_-(x;X)$ designates the variation of the
distribution due to the
breaking of some contacts. We can formally write it as
\begin{equation}
\Delta Q_-(x;X) = P(x) \, \Delta \! X \, Q(x;X)  \; ,
\label{Q3}
\end{equation}
where $P(x) \, \Delta \! X$ is the fraction of contacts displaced by
$x$ from their fully relaxed state
which break when their displacement is increased by $\Delta X$. This
fraction is related to the individual properties of the contacts,
which are determined by the distribution of static thresholds
of each contact $P_c(x)$ defined in Sec.~\ref{earthquakes}.
According to the definition of  $P_c(x)$, the total number of unbroken
contacts when the stretching of the asperities is equal to $x$ is given
by $N_c \int_x^{\infty} P_c(\xi) \; d \xi$. The contacts that break
when their displacements are increased by $\Delta \! X$ are those which
have their thresholds between $x$ and $x + \Delta \! X$, i.e.\ $N_c
P_c(X) \, \Delta \! X$. Therefore
\begin{equation}
P(x) = \frac{P_c (x)}{\int_{x}^{\infty}  P_c (\xi) \; d\xi} \; .
\label{PcP}
\end{equation}

The broken contacts relax and have to be added to the distribution
around $x \sim 0$, leading to the third term in Eq.~(\ref{Q2}).
We denote by $R(x)$ the normalized distribution of the stretching of the
asperities that form a new contact after a slip event.
Writing that all broken contacts described by $\Delta Q_-(x;X)$
immediately reappear with the distribution $R(x)$
(see Appendix~\ref{delay}), we get
\begin{equation}
\Delta Q_+(x;X) = R(x) \int_{-\infty}^{\infty}
dx^{\prime} \, \Delta Q_-(x^{\prime};X).
\label{Q4}
\end{equation}

Equation (\ref{Q2}) can be rewritten as
\[ \left[ Q(x;X + \Delta \! X) - Q(x;X) \right] +
\left[ Q(x;X) - Q(x - \Delta \! X;X) \right] \]
\[ =  - \Delta Q_-(x;X) + \Delta Q_+(x;X). \]
Taking the limit $\Delta \! X \to 0$, we finally get the
integro-differential equation
\begin{eqnarray}
\frac{ \partial Q(x;X) }{ \partial x} +
\frac{ \partial Q(x;X) }{ \partial X} +
P(x) \, Q(x;X)
\nonumber
\\
= R(x) \int_{-\infty}^{\infty} dx^{\prime} \, P(x^{\prime}) \, Q(x^{\prime};X),
\label{Q5a}
\end{eqnarray}
which has to be solved with the initial condition
\begin{equation}
Q(x;0) = Q_{\rm ini} (x).
\label{Q5b}
\end{equation}
Note that  $Q_{\rm ini} (x)$ cannot be an arbitrary function,
because the contacts that exceed their stability threshold, must relax
from the very beginning.

In the following we select $R(x)=\delta(x)$, i.e.\ we assume that
a broken contact
sticks again only after a complete relaxation. This is physically
reasonable and simplifies the
analytical calculations, although this restriction could be easily
lifted for numerical solutions of Eq.~(\ref{Q5a}).

Integrating both sides of Eq.~(\ref{Q5a}) over $x$ from $-\infty$ to $\infty$,
we see that the normalization condition (\ref{Q5c}) is satisfied,
taking into account $Q(-\infty,X)=Q(+\infty,X)=0$
because contacts cannot be infinitely stretched without breaking.

Once the distribution $Q(x;X)$ is known, we can calculate the total force $F$
with Eq.~(\ref{Q6})
and find the static friction force as the maximum of $F(X)$, i.e.,
$F_s = F(X_s)$,
where $X_s$ is a solution of the equation
\begin{equation}
F^{\prime}(X) \equiv dF(X)/dX =0.
\label{Q6b}
\end{equation}

%-------------------------------------------------------------------
\subsection{Steady state solution}

The smooth-sliding solution $Q_s (x)$, i.e. the solution which does
not depend on $X$,
can easily be found directly
% and corresponds to the solution derived for the EQ model.
\cite{P1995}.
In the steady state Eq.~(\ref{Q5a}) reduces to
\begin{equation}
\frac{d Q(x) }{d x} +
P(x) \, Q(x) = \delta(x) \int_{0}^{\infty} dx^{\prime} \,
P(x^{\prime}) \, Q(x^{\prime}).
\label{Q10}
\end{equation}
In setting the lower bound of the integral we have assumed that
$P_c(x)=0$ for $x \leq 0$, which agrees with its physical meaning
because, if $x < 0$, a positive variation $\Delta X$ actually reduces
the absolute value of the force on a contact
so that it does not cause its breaking.

Its general solution can easily be derived for
$x > 0$ and leads to
\begin{equation}
Q_s (x) = C^{-1} \Theta(x) E_P (x),
\label{Q11a}
\end{equation}
where $\Theta(x)$ is the Heaviside step function,
$\Theta(x)=0$ for $x<0$ and
$\Theta(x)=1$ for $x \geq 0$,
\begin{equation}
E_P (x) = e^{-U(x)},
\;\;\;
U(x) = \int_{0}^{x} d\xi \, P(\xi) \; ,
\label{Nik04}
\end{equation}
and this solution also verifies the equation in the limit $x \to 0$.
The normalization condition for $Q_s(x)$ gives
\begin{equation}
C = \int_{0}^{\infty} dx \, E_P (x).
\label{Q11c}
\end{equation}

The friction force is then equal to
\begin{equation}
F = (N_c \langle k_i \rangle /C) \int_{0}^{\infty} dx \, x E_P (x).
\label{Q11d}
\end{equation}

Two simple examples are considered in Appendix~\ref{simple}.

%-------------------------------------------------------------------
\subsection{Relation between the EQ and ME}

The EQ model is expressed in terms of the distribution of breaking
thresholds of the contacts $P_c(x)$. Therefore, to connect the two
approaches it is interesting to relate the steady state solution of
the ME equation and the function $P_c (x)$.
Using Eq.~(\ref{PcP}) we get
\begin{equation}
  Q_s (x) = C^{-1} \Theta(x) \exp \left[ - \int_{0}^{x} d\xi \,
\frac{P_c (\xi)}{\int_{\xi}^{\infty} P_c (\xi') \; d\xi'} \right] \; .
\end{equation}
The integration in the exponential can be readily performed since the
integrand is of the form $(d/d\xi) \ln \int_{\xi}^{\infty} P_c(\xi')
\, d\xi'$. We get
\begin{equation}
   Q_s (x) = C^{-1} \Theta(x) \int_{x}^{\infty}  P_c (\xi) \; d\xi ,
\end{equation}
or ($x>0$)
\begin{equation}
  \frac{dQ_s(x)}{dx} = - C^{-1} P_c(x)
\end{equation}
in agreement with the observations deduced from the numerical
simulations of the EQ model (Fig.~\ref{A02c}, lower panel).
For the parameters used in Figs.~\ref{A02a}--\ref{A02c}, 
the numerical constant is  $C=0.9999$.

Notice also the useful relationships $U^{\prime} (x)=P(x)$,
$E_P^{\prime} (x) = -E_P (x) P(x)$,
$E_P (x)=1$ for $x \leq 0$, and
\begin{equation}
P_c (x) = P(x) \, E_P (x) \;\;\; {\rm for} \;\;\; x>0,
\label{PcPx2}
\end{equation}
the latter may be used in numerical solution of the ME instead of
Eq.~(\ref{PcP}).

For the simple example of a rectangular $P_c (x)$ distribution,
which is considered in Appendix~\ref{simple},
the model admits an exact solution both for the EQ and ME approaches.
We checked that the earthquake model with the distribution (\ref{Q22a})
and the master equation model with the expression (\ref{Q22b}) for $P(x)$
exactly have the same solution for any initial configuration.

%-----------------------------------------------------------------
\subsection{Non-stationary solution}
\label{time-dependent}

\begin{figure}[h] %[t] \bigskip
\includegraphics[width=8cm]{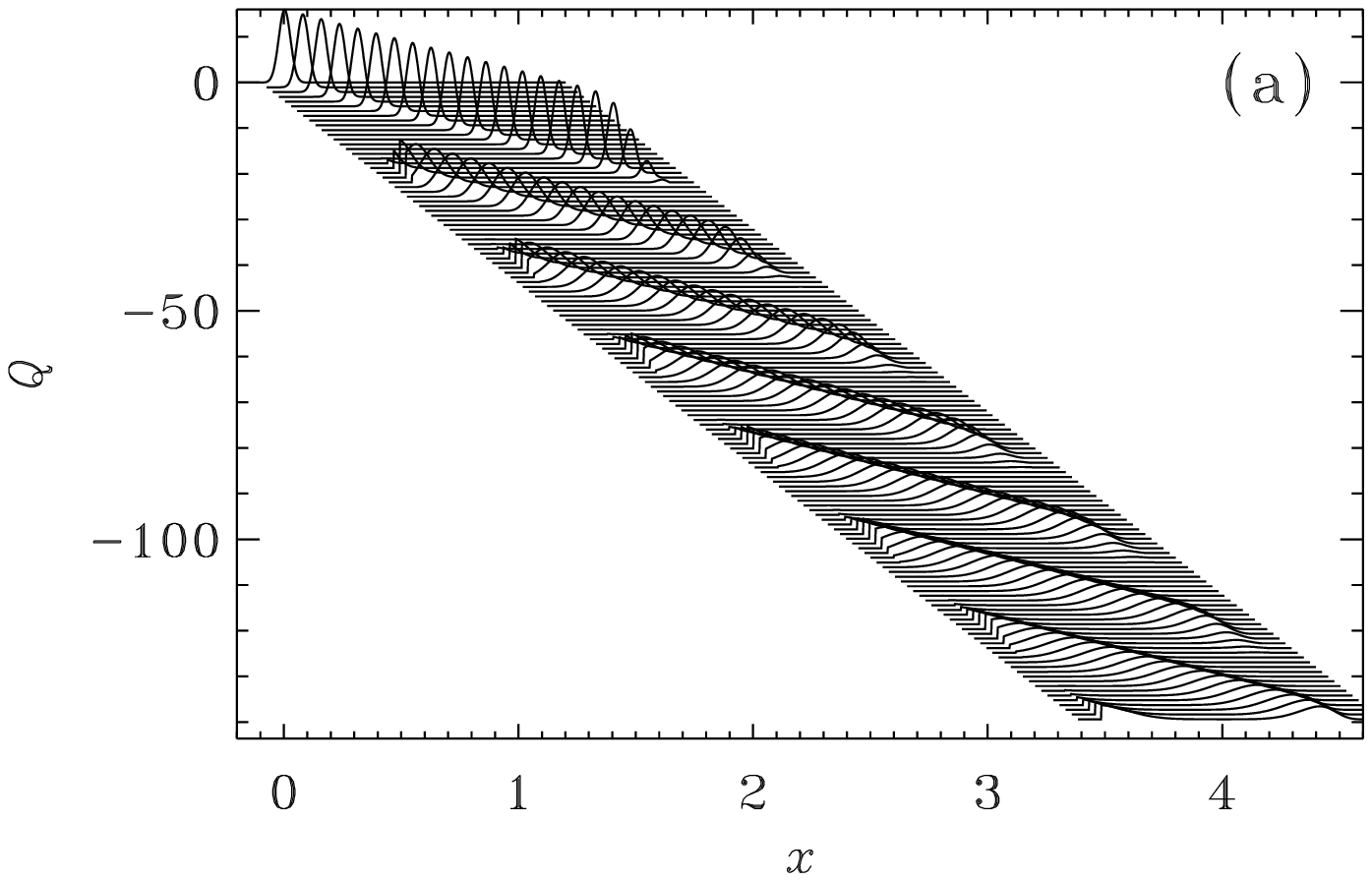} % clip
\includegraphics[width=7.8cm]{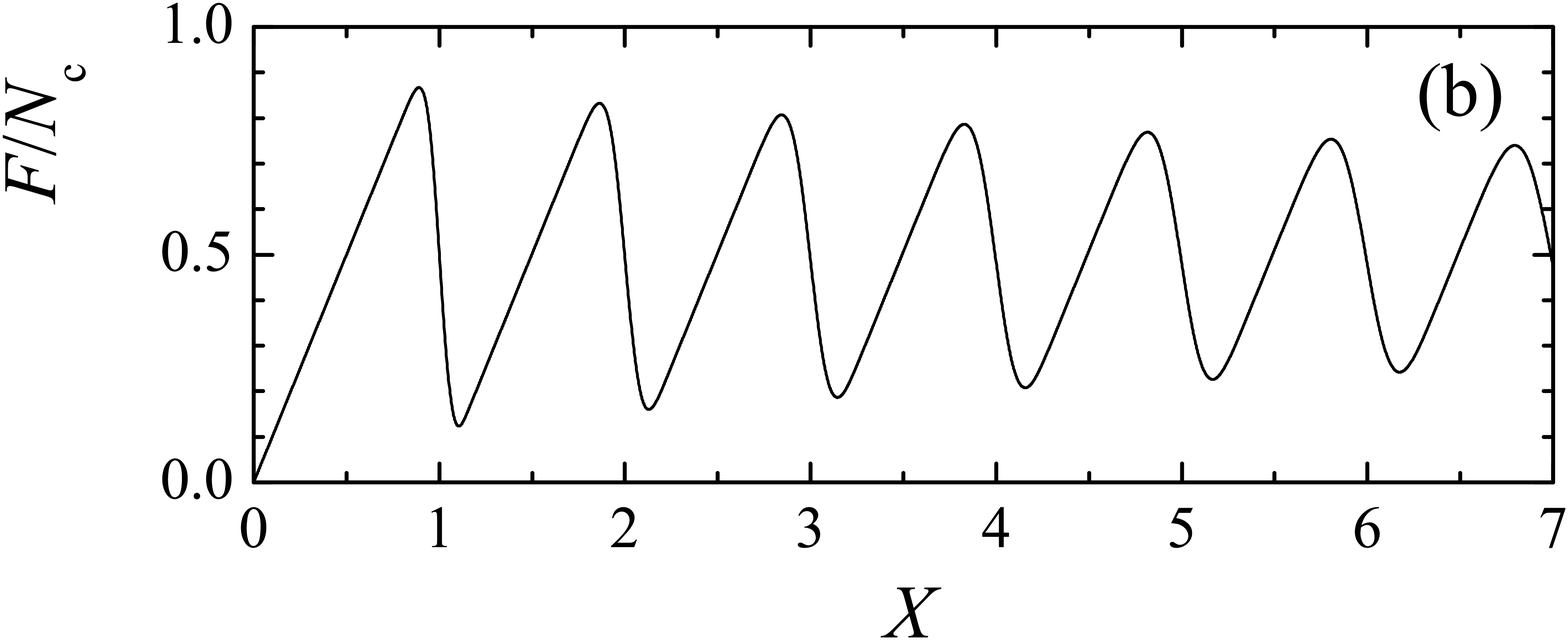}
\caption{\label{A03a}
(a) Solution of the master equation in the initial stage when $X$
  starts to grow.
The model parameters are the same as in Fig.~\ref{A02a}.
% (dotted curves show the distribution $P_c (f)$).
(b) The corresponding dependence $F(X)$.
%[use A03/fig5a.pro = A03a + master-short.eps]
}
\end{figure}

\begin{figure}[h] %[t] \bigskip
\includegraphics[width=8cm]{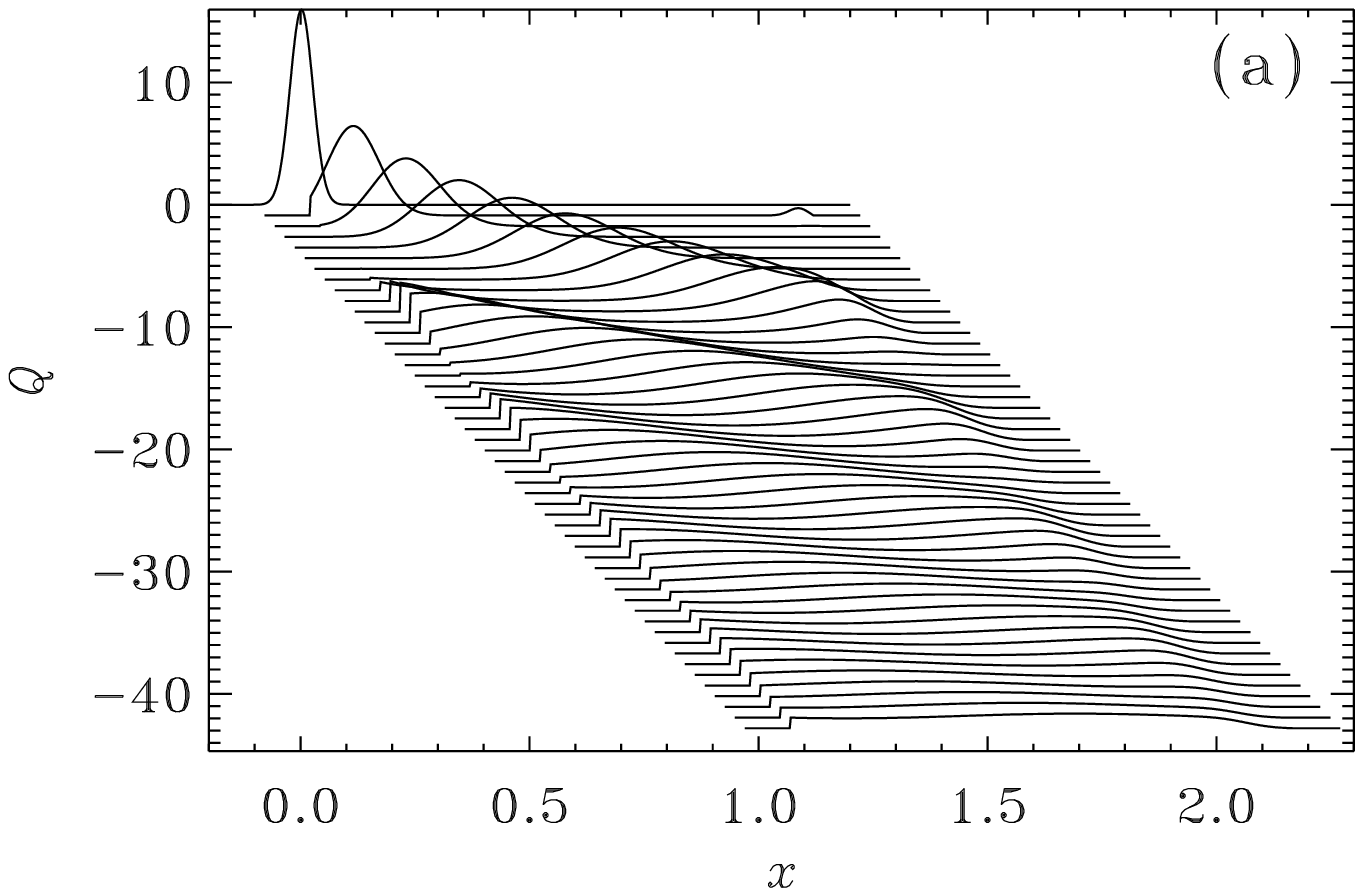} % clip
\includegraphics[width=8.2cm]{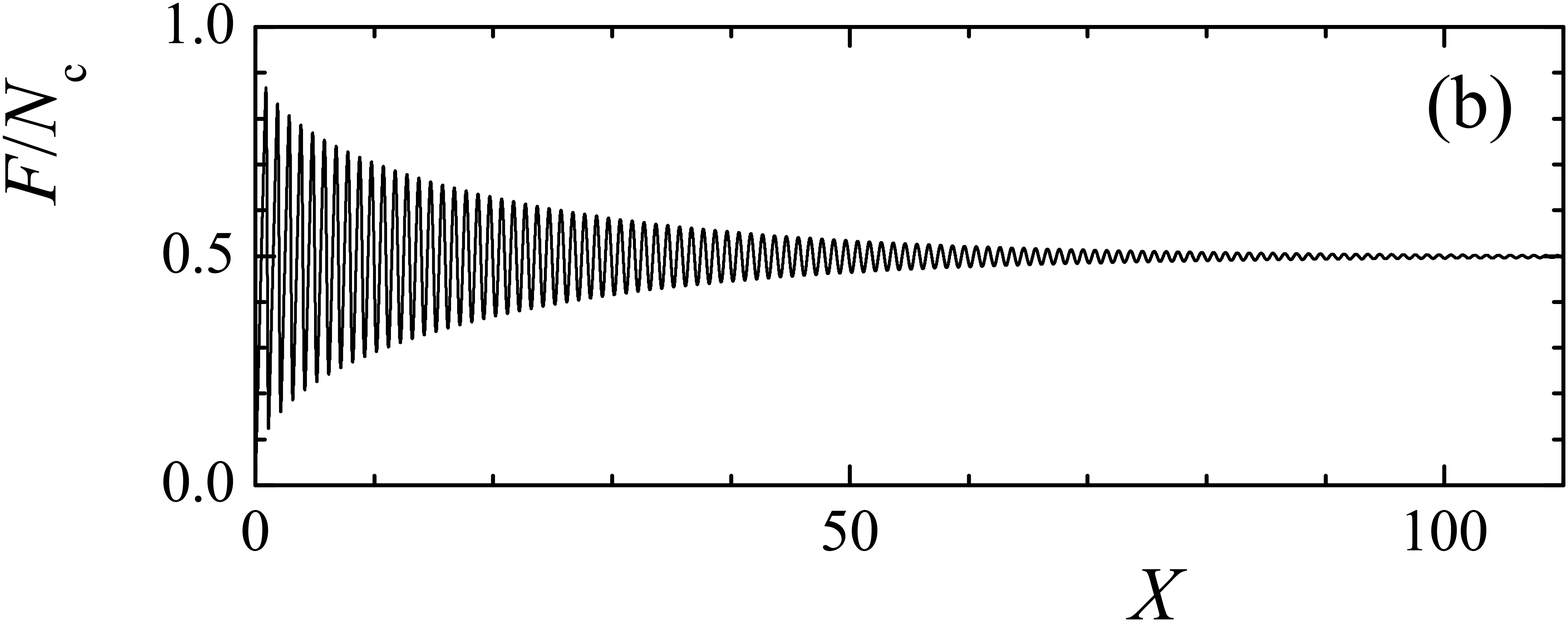}
\caption{\label{A03b}
The same as in Fig.~\ref{A03a} for long times
with the increment $\Delta X =1.09$.
%[use fig2bMP PRL = A03b + master-long.eps]
}
\end{figure}

The numerical solution of the master equation (\ref{Q5a})
when $X$ starts to grow due to an external driving
is presented in Figs.~\ref{A03a} and~\ref{A03b}
for a Gaussian distribution of static contact thresholds,
$P_c (x) = G(x; \bar{x}_s, \sigma_s)$.
\begin{figure}[h] %[t] \bigskip
\includegraphics[clip, width=8cm]{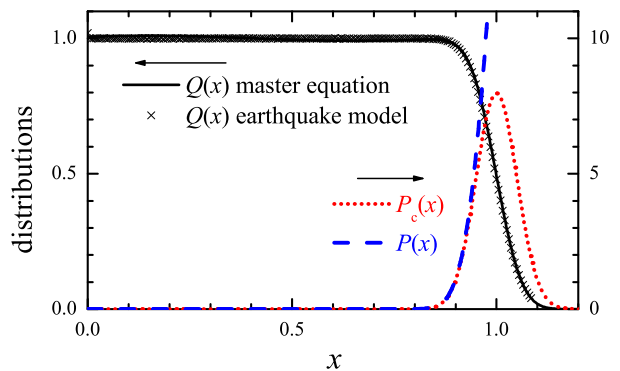}
\caption{\label{A03c}(Color online):
The final distribution $Q(x)$ for the parameters used in Fig.~\ref{A03b}
(solid curve;
crosses show the averaged final distribution for the earthquake model
from Fig.~\ref{A02c}).
The red dotted curve shows the distribution $P_c (x)$,
blue broken curve -- the function $P(x)$.
%[A03c; use A03.opj]
}
\end{figure}
This solution may be compared with that of the EQ model of
Figs.~\ref{A02a} and~\ref{A02b} for the same initial condition.
One can see that they are almost identical,
except for the noise on the earthquake model distributions.
The distribution $Q(x;X)$ always approaches the
stationary distribution $Q_s (x)$ given by Eq.~(\ref{Q11a}).
The final distributions of the EQ model and the ME approach
are compared in Fig.~\ref{A03c}.

%------------------------------------------------------------------
\smallskip
Let the distribution $P_c (x)$
be characterized by the average value $\bar{x}_s$ and the dispersion $\sigma_s$.
Studying the evolution presented in Fig.~\ref{A03a}, we observe that
every-time $X$ increases by $\bar{x}_s$ the distribution  $Q(x;X)$
broadens so
that its standard deviation grows by $\sim \sigma_s$.
Therefore, any initially peaked distribution
tends to approach the stationary one according to
$| Q(x;X) - Q_s (x) | \propto \exp (-X/X^*)$, where
$X^* \sim \bar{x}_s^2 / \sigma_s$.

\smallskip
For one particular but important choice of the initial distribution,
when all contacts are relaxed at the beginning, $Q_{\rm ini} (x) = \delta (x)$,
one can analytically express the initial evolution
of the solution versus $X$.
Namely, if $P(x)=0$ for the stretchings $x < x_c^*$, then, as can be
checked by its substitution in the master equation, 
the solution for
the displacement $X < 2 x_c^*$ is the following:
at the beginning, for the displacement $0 < X < x_c^*$, the solution is trivial,
$Q(x;X) = \delta(x-X)$ and $F(X) = N_c \langle k_i \rangle X$;
for larger displacements, $x_c^* < X < 2 x_c^*$, the solution is
\begin{equation}
Q(x;X) = \left\{
\begin{array}{ll}
\Theta(x) \, b(X-x) & {\rm for} \;\;\; 0 \leq x < X - x_c^*,
\\
0 & {\rm for} \;\;\; X - x_c^* \leq x < x_c^*,
\\
a(X) \, \delta(X-x) & {\rm for} \;\;\; x \geq x_c^*,
\end{array}
\right.
\label{Q17}
\end{equation}
where
\begin{equation}
b(\xi) = P(\xi) \, a(\xi)
\label{Q18}
\end{equation}
and $a^{\prime} (\xi) = - P(\xi) a(\xi)$, or
\begin{equation}
a(\zeta) = \exp \left[ -\int_{x_c^*}^{\zeta} d\xi \, P(\xi) \right],
\;\;\; \zeta \geq x_c^*.
\label{Q19}
\end{equation}
Then one can calculate the friction force for the displacement
interval $x_c^* < X < 2 x_c^*$
\begin{equation}
F(X) = N_c \langle k_i \rangle \left[ X a(X) + (X-x_c^*) -
  \int_{x_c^*}^{X} d\xi \, a(\xi) \right].
\label{Q20}
\end{equation}

The solution (\ref{Q20}) allows us to find the static friction force
$F_s = F(X_s)$ with Eq.~(\ref{Q6b}), which reduces to the equation
$X_s a(X_s) P(X_s) =1$
and corresponds to the first maximum on the variation of $F(X)$.
As was pointed out by Farkas \textit{et al.}~\cite{FDW2005},
the ratio $F_s /F_k$, where $F_k$ is the kinetic friction, i.e.\ the
plateau reached by $F(X)$, takes values between 1 and~2
[larger values corresponding to narrower distributions $P_c (x)$],
and it is determined solely by the initial distribution $Q_{\rm ini} (x)$,
reaching a maximum for $Q_{\rm ini} (x) = \delta(x)$
(it was called ``the initial coherence in strain distribution of the contacts''
in Ref.~\cite{FDW2005}).

%------------------------------------------------------------------
\medskip
Thus, in a general case, as $X$ grows the solution of the master
equation always approaches the smooth-sliding state given by
Eqs.~(\ref{Q11a}--\ref{Q11c}).
However, there is one exception to this general scenario,
when the model admits a periodic solution.
This is the {\it singular\/} case
when all contacts are identical, i.e., all contacts are characterized by
the same static threshold $x_s$,
so that $P_c(x) = \delta (x-x_s)$. % , or $P(f)=0$ for $f<f_s$ and
                                % $P(f)=\infty$ for $f \geq f_s$.
The steady-state solution of Eq.~(\ref{Q5a}) for this case is
described in Appendix~\ref{singular}.
We emphasize that the solution (\ref{Q17}) is valid for the \textit{continuous}
distribution $P_c(x)$ only,
it cannot be used for the singular $P_c(x) = \delta (x-x_s)$ case.

\subsection{Nonrigid substrates: stick-slip}
\label{nonregid}

The master equation allows us to compute the friction force $F(X)$
when the bottom of the sliding block is displaced by $X$. But
actually in an experiment one does not control $X$. The displacement
is caused by a shearing force $F_T$ applied to the top of the sliding
block, which displaces its top surface by $X_T$. As the strain on the
sliding block is usually small, its deformation can be assumed to be
elastic, so that $X_T$ is related to the applied force by
\begin{equation}
F_T=K(X_T -X),
\label{Q16x2}
\end{equation}
where $K$ is the shear elastic constant of the solid block. The
solution that we discussed above amounts to assuming that the sliding
block was infinitely rigid, $K \to \infty$, so that $X_T = X$ at any
time. In this case we found that the sliding always tends to a
smooth-sliding  steady state, as expected for a stiff system
\cite{P0,BC2006,BN2006}.
Let us now consider the case of finite $K < \infty$. The total force applied
to the bottom of the sliding block, which determines its displacement
$X$, is the sum of the applied force and the friction force
\begin{equation}
F_{\mathrm{tot}} = K(X_T - X) - F(X) \; .
\end{equation}
It can be viewed as derived from the potential
\begin{equation}
  V(X_T,X) = \frac{1}{2} K (X_T - X)^2 + \int_0^X F(\xi) \, d\xi \; ,
\end{equation}
which determines the behavior of the sliding block subjected to
friction and the applied force.

\smallskip
A necessary condition for smooth sliding is that $X_T$ and $X$ grow
together, with $X_T - X = B$, where $B$ is a constant that measures
the shear strain in the sliding block. This leads to the condition
\begin{equation}
  \frac{\partial V}{\partial (X_T - X)} = \frac{\partial V}{\partial X} = 0
  \; ,
\end{equation}
which simply means that the total force on the sliding interface
vanishes. Smooth sliding also requires this state to be stable
\begin{equation}
  \frac{\partial^2 V}{\partial (X_T - X)^2} = \frac{\partial^2 V}{\partial
    X^2} \ge 0 \quad \mathrm{or} \quad F'(X) \ge -K \; .
\label{Q16}
\end{equation}
These arguments exactly coincide with the ``elastic instability''
widely discussed in literature (e.g., see Ref.~\cite{BC2006} and
references therein).

If we start from relaxed asperities, in the early stage of the motion
$F(X)$ is a growing function of $X$, and then it passes by a maximum
when some contacts start to break and reform at lower asperity stress.
Depending on the value of $K$, two situations are possible. For large
$K$ (stiff block), $F'(X)$ never falls below $-K$ and the smooth
sliding is a stable steady state. In this case the system
evolves towards the regime $F(X)=$~Const defined by Eq.~(\ref{Q11d}).
For small $K$ (soft block), $F'(X)$
can become smaller than $-K$ and the stability condition~(\ref{Q16})
is no longer valid. In this case Eq.~(\ref{Q16}),
in the the limiting case of the equality,
defines the maximal displacement $X_m$ that the contacts can sustain.
A larger displacement breaks all the contacts simultaneously, causing
a quick slip of the block before the contacts form again with a fresh
distribution $Q(x,X_m)=R(x)$ and the same process can repeat again.
The system may reach the regime of stick-slip periodic motion~\cite{BT2009}.
Using the solution~(\ref{Q20}) for the delta-function initial distribution,
we get
$F^{\prime} (X) = N_c \langle k_i \rangle \left[ 1 - X P(X) \, a(X) \right]$,
so that we can find the period $X_{ss}$ of the
stick-slip motion with Eq.~(\ref{Q16}).
Ignoring the slip time, it is given by the solution of the equation
$X \, a(X) P(X) = 1 +K/N_c \langle k_i \rangle$ for $X > x_c^*$.
Note that in the case of an extremely soft substrate, $K \to 0$,
the motion (almost always) corresponds to stick-slip,
as it should for such a soft system
\cite{P0,BC2006,BN2006}.

Thus, depending on the distribution $P_c (x)$, the system demonstrates either
stick-slip or smooth sliding.
Smooth sliding is achieved if $K > K^*$, otherwise the elastic
instability occurs
which may result in stick-slip.
As $K^* = \max [- F'(X)] $ and $F(X)$ reaches its maximum for
$X \simeq \bar{x}_s$,
one can estimate that $K^* \sim N_c \langle k \rangle \, \bar{x}_s/\sigma_s$;
thus, the ratio $\sigma_s/\bar{x}_s$ controls the appearance of the
elastic instability.

%------------------------------------------------------------------
\section{Temperature effects}
\label{temperature}

The effect of a nonzero temperature is connected with a change
of the fraction density of breaking contacts $P(x)$
in the master equation (\ref{Q5a})
as first discussed by Persson~\cite{P1995}.
Indeed, for a single contact with the static threshold $x_s$ at zero
temperature, the contact does not break at all for $x<x_s$.
But when $T>0$, the contact may relax due to a thermally activated jump
before the threshold $x_s$ is reached.
The rate $h(x;x_s)$ of this process is defined by
\begin{equation}
dQ(t)/dt = h(x;x_s) \, Q(t), \;\;\; x<x_s \; ,
\label{temper1}
\end{equation}
where $Q(t)$ is the probability that a contact existing at $t=0$ is
not thermally broken at time $t$.
For a set of contacts, Eq.~(\ref{temper1}) has to be generalized to
\begin{equation}
dQ(t)/dt = H(x) \, Q(t)
\label{temper2}
\end{equation}
with
\begin{equation}
H(x) =\int_{x}^{\infty} dx^{\prime} \, h(x;x^{\prime}) \, P_c (x^{\prime}).
\label{temper3}
\end{equation}
For a sliding at velocity $v$ so that $X=vt$,
the thermally activated jumps can be incorporated in the master equation,
if we use
a corrected expression $P_T (x)$ defined by
\begin{equation}
P_T (x) = P(x) + H(x)/v \; ,
\label{temper4}
\end{equation}
instead of the zero-temperature breaking fraction density $P(x)$.

The rate of thermal activation of the contacts, $h(x;x_s)$ in
Eq.~(\ref{temper1}),
can  be estimated by the Kramers relation.
Let $V(x)$ be the binding energy of the contact,
and $\Delta E(x;x_s) = V(x_s)-V(x)$, % be
the activation energy for the contact to break.
For ``soft'', or ``weak'' contacts, when $\Delta E(0;x_s) \agt k_B T$,
the rate $h(x;x_s)$ is given by \cite{P1995,FKU2004}
\begin{equation}
h(x;x_s) \approx \omega \, \exp \left[ -\Delta E(x;x_s)/k_B T \right],
\label{temper5}
\end{equation}
where $\omega$ is a prefactor corresponding to the attempt frequency.
For an overdamped dynamics of the contacts
$\omega \approx \omega_0^2 /2\pi \eta$
with the characteristic frequency $\omega_0^2 \sim k/m \sim c^2/A$
($m$ is the contact mass)
and the damping coefficient $\eta \sim c/\sqrt{A}$
which gives $\omega \sim c/2\pi\sqrt{A} \sim 10^{10}$~s$^{-1}$ \cite{P1995}.
If we assume that the binding potential of a contact can be
approximated by the elastic properties of this contact, then
$ V(x)={1 \over 2} \, kx^2%  = {1 \over 2} \, (kx)^2 /k = f^2 /2k
$. % \label{temper7}  \end{equation}
In this case the activation energy for contact breaking takes the form
\cite{P1995}
\begin{equation}
\Delta E(x;x_s) = {1 \over 2} \, k(x_s^2 -x^2) \approx k x_s \, (x_s -x),
\label{temper8}
\end{equation}
and the function $H(x)$ in Eq.~(\ref{temper4}) is given by
\begin{equation}
H(x)=\omega \, e^{k x^2 /2k_B T} \int_{x}^{\infty} d\xi \, P_c (\xi)
\, e^{-k \xi^2 /2k_B T}.
\label{temper11}
\end{equation}

\begin{figure} [h] %[t] \bigskip
\includegraphics[clip,width=8cm]{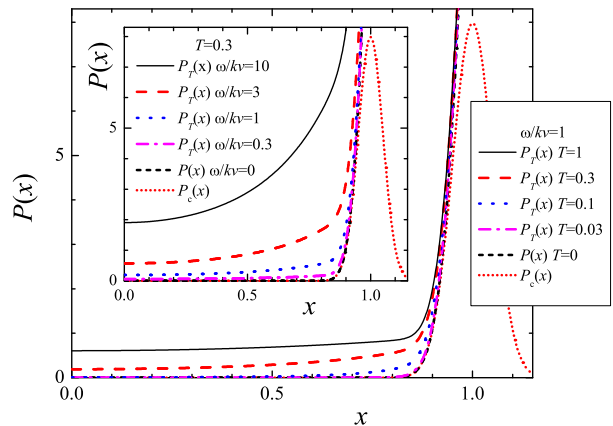} % clip
\caption{\label{A08a} (Color online):
The rate $P_T (x)$ for soft contacts at different temperatures $T=0$
(black short dashed), 
0.03 (magenta dash-dotted), 0.1 (blue dotted), 0.3 (red dashed) and~1
(black solid curve) 
for a given velocity, $\omega /kv=1$.
Inset: the rate $P_T (x)$ at different velocities $\omega /kv=0$
(black short dashed), 
0.3 (magenta dash-dotted), 1 (blue dotted), 3 (red dashed) and~10
(black solid curve) 
for a fixed temperature $T=0.3$.
The red short-dotted curve shows the distribution $P_c (x)$
(Gaussian with $\bar{x}_s =1$ and $\sigma_s =0.05$).
%[use Rotate/idl/earth12c.pro and Rotate/idl/T.gt.0/F(X)-1.opj,
%  F(X)-2.opj = A08a]
}
\end{figure}

For ``stiff'', or ``strong'' contacts, when $\Delta E(0;x_s) \gg k_B
T$, contact breaking only occurs with a significant probability when
the stretching $x$ of a  contact is close to the threshold $x_s$. The
harmonic expression of the binding energy can no longer be used and a
cubic approximation is more appropriate. This leads to a correction
in the activation energy \cite{Garg1995,Dudko2003}
\begin{equation}
\Delta E(x;x_s) = \Delta E(0;x_s) \left( 1-x/x_s \right)^{3/2},
\label{temper9}
\end{equation}
and a renormalization of the prefactor $\omega$,
\begin{equation}
\omega \to \omega \, \left( 1-x/x_s \right)^{1/2},
\label{temper10}
\end{equation}
so that the function $H(x)$ is given by
\begin{equation}
H(x)=\omega \, \int_{x}^{\infty} d\xi \, P_c (\xi)
\left( 1-\frac{x}{\xi} \right)^{\frac{1}{2}}
\exp \left[ -\frac{k \xi^2 (1-x/ \xi)^{\frac{3}{2}} }{2k_B T} \right].
\label{temper12}
\end{equation}

The contributions (\ref{temper11}) or (\ref{temper12}) to the rate $P_T (x)$
lead to appearance of a low-$x$ tail.
Its magnitude grows with temperature as well as with a decrease
of the driving velocity $v$ as demonstrated in Fig.~\ref{A08a}.

\begin{figure} % [h] %[t] \bigskip
\includegraphics[clip,width=8cm]{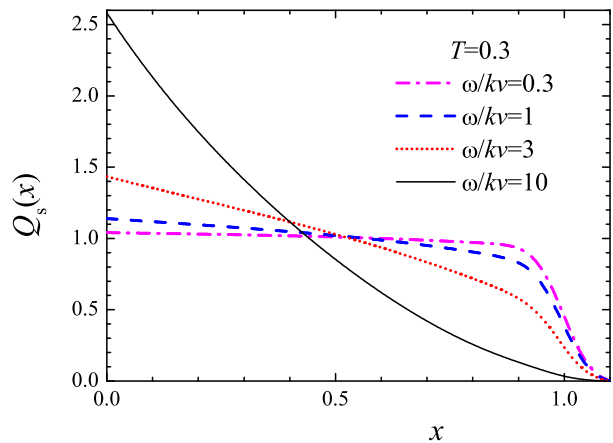} % clip
\caption{\label{A08b} (Color online):
The steady-state distribution $Q_s (x)$ for $T=0.3$ and
different velocities $\omega /kv=0.3$ (magenta dash-dotted),
1 (blue dashed),
3 (red short-dotted)
and~10 (black solid curve) for the soft contacts.
The distribution $P_c (x)$ is
Gaussian with $\bar{x}_s =1$ and $\sigma_s =0.05$.
%[use Rotate/idl/earth12c.pro and Rotate/idl/T.gt.0/F(X)-2.opj = A08b]
}
\end{figure}

Raising temperature leads to a shift of the distribution $Q(x)$ to
lower values,
so that the friction force decreases when $T$ grows.
This effect is larger for lower sliding velocities
as shown in Fig.~\ref{A08b}.
In the limit $v \to 0$, all contacts finally break if $T \neq 0$,
so that $Q_s (x) \to \delta (x)$ and $F_k \to 0$;
in this limit we have a smooth sliding associated to a creep motion of
the contacts.

\begin{figure} % [h] %[t] \bigskip
\includegraphics[clip,width=8cm]{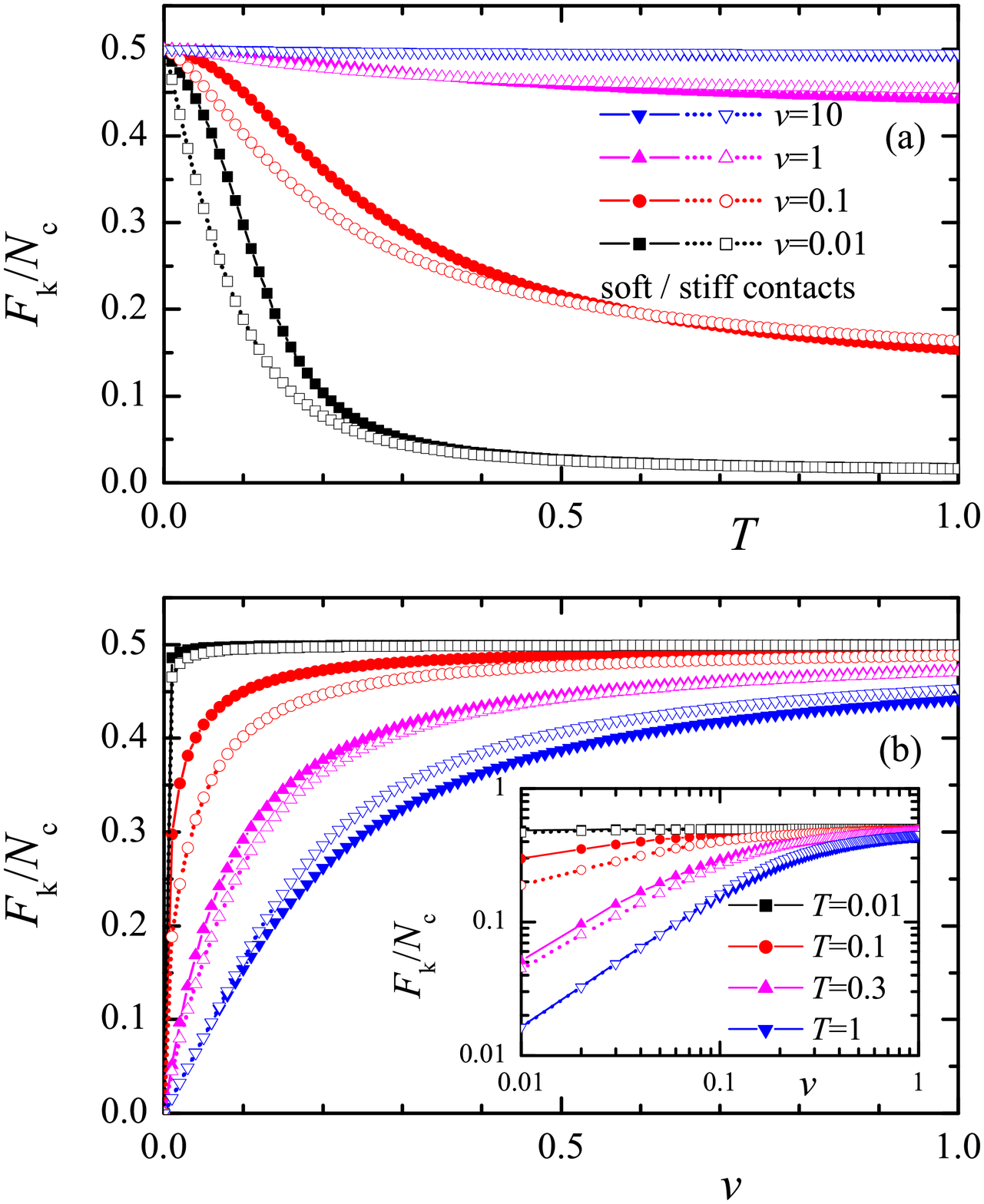} % clip
\caption{\label{A08c} (Color online):
(a) Dependence of the kinetic friction force $F_k$ in the steady state
on the temperature $T$ for different velocities $v=0.01$ (black squares),
0.1 (red circles),
1 (magenta up triangles)
and~10 (blue down triangles).
(b)~$F_k (v)$ for different temperatures $T=0.01$ (black squares),
0.1 (red circles),
0.3 (magenta up triangles)
and~1 (blue down triangles);
the inset shows the same in log-log scale.
Solid curves and symbols are for soft contacts,
and broken curves and open symbols, for stiff contacts.
The distribution $P_c (x)$ is
Gaussian with $\bar{x}_s =1$ and $\sigma_s =0.05$;
$\omega /k=1$.
%[use Rotate/idl/earth12c.pro, earth12d.pro and
%Rotate/idl/T.gt.0/F(Tv).opj, F(Tv)-stiff.opj = A08c]
}
\end{figure}

The variation of $F_k (T)$ versus $T$
for different sliding velocities in the smooth sliding regime
for soft and stiff contacts is presented in Fig.~\ref{A08c}a.
As expected, the friction force decreases when $T$ grows and tends to
zero when $T \to \infty$.
Figure~\ref{A08c}b shows the dependence of kinetic friction
on the sliding velocity.
The force $F_k$ monotonically increases with $v$, approaching the $T=0$ limit
when $v \to \infty$.
The dependence $F_k (v)$ at smooth sliding was estimated analytically
by Persson \cite{P1995}
(see also Ref.~\cite{SW2009}) who found
$F_k (v) \propto v$ in the low-velocity limit,
$F_k (v) \propto \ln v$ for intermediate velocities, and
$F_k (v) - F(\infty) \propto -\omega T^2 /kv$ in the high-velocity case.

Thus, for $T>0$ the friction force $F_k$ increases with the driving
velocity, i.e.\ $dF_k (v)/dv >0$, which stabilizes the smooth sliding regime.
However, experiments show that the temperature-induced
velocity-strengthening only dominates
the aging-induced velocity-softening (see Sec.~\ref{aging}) at
high velocities
$v > 10 - 100$~$\mu$m/s~\cite{BC2006}.

\begin{figure} % [h] %[t] \bigskip
\includegraphics[clip,width=8cm]{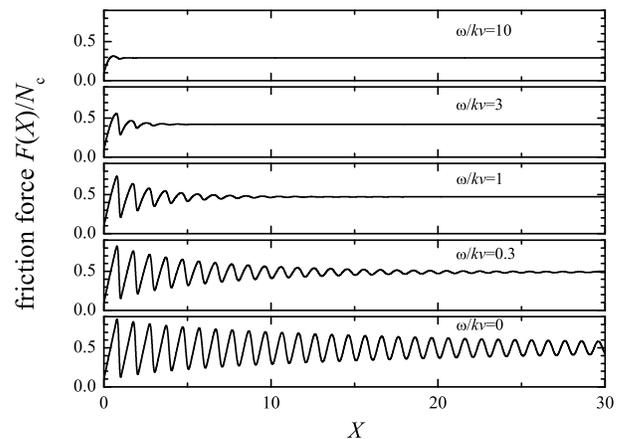} % clip
\caption{\label{A08d}
The dependences $F(X)$ for $T=0.3$ at different driving velocities $v$,
$\omega /kv=0$, 0.3, 1, 3 and~10.
The distribution $P_c (x)$ is Gaussian with
$\bar{x}_s =1$ and $\sigma_s =0.05$,
the initial distribution $Q_{\rm ini} (x)$ is
Gaussian with $\bar{x}_{\rm ini} =0.1$ and $\sigma_{\rm ini} =0.025$.
%[use Rotate/idl/earth12b.pro and Rotate/idl/T.gt.0F(X)-1.opj,
%  F(X)-2.opj = A08d]
}
\end{figure}

\begin{figure} % [h] %[t] \bigskip
\includegraphics[clip,width=8cm]{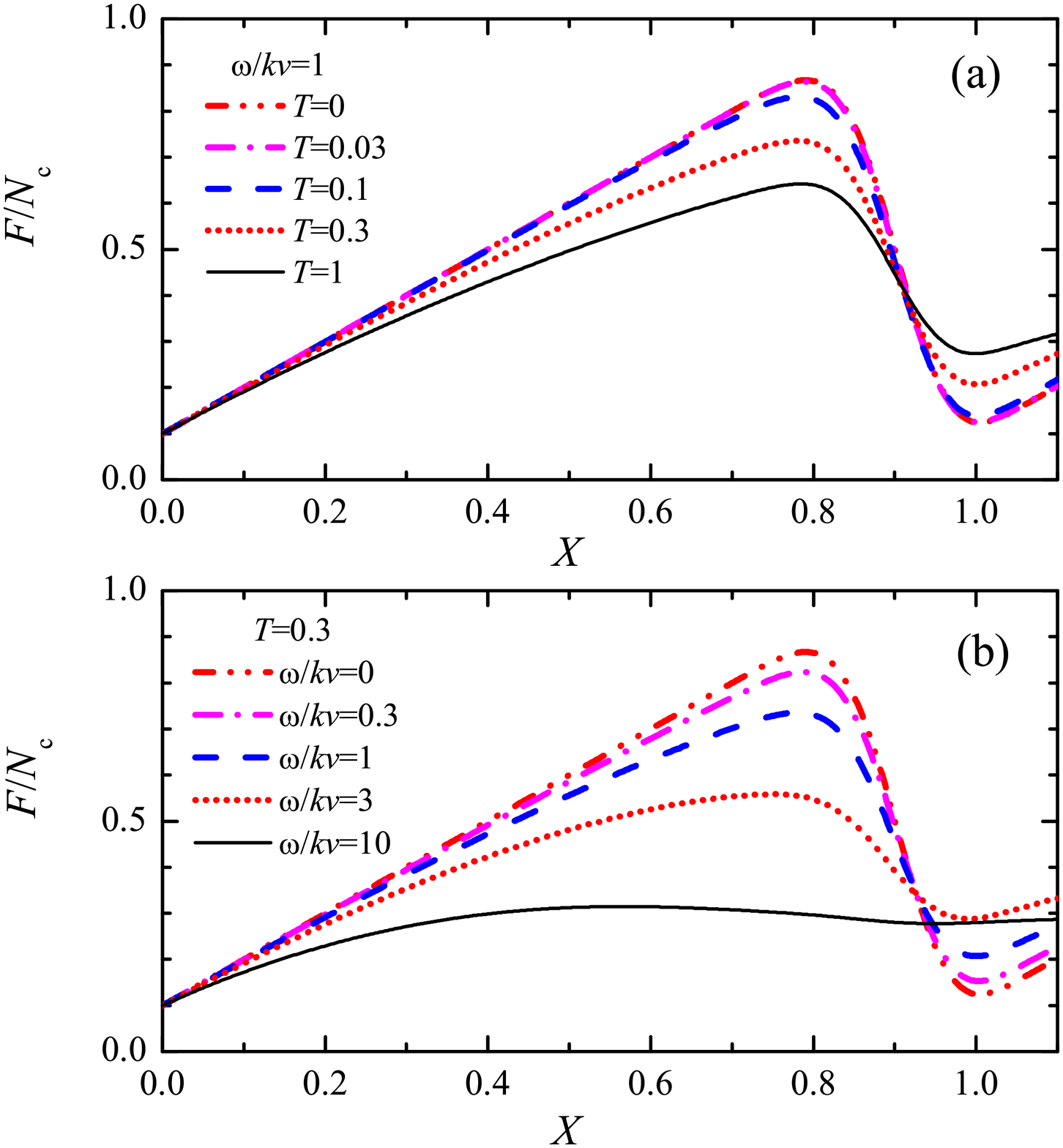} % clip
\caption{\label{A08e} (Color online):
The same as in Fig.~\ref{A08d} for short~$X$.
(a) is for different temperatures $T=0$ (red dash-dot-dotted),
0.03 (magenta dash-dotted),
0.1 (blue dashed),
0.3 (red dotted)
and~1 (black solid curve)
at the fixed velocity, $\omega /kv=1$;
(b) is for different velocities $v$, $\omega /kv=0$ (red dash-dot-dotted),
0.3 (magenta dash-dotted),
1 (blue dashed),
3 (red dotted)
and~10 (black solid curve)
at the fixed temperature $T=0.3$.
%[use A08e]
}
\end{figure}

Of course, a nonzero temperature also affects the dynamics
of the approach to the steady state as shown in Fig.~\ref{A08d} for
different driving velocities.
The higher the temperature and/or the lower the velocity are,
the lower is the static friction force $F_s$
[determined by the first maximum of the $F(X)$ dependence], and
the faster $F(X)$ approaches the steady-state smooth sliding.
It is important to consider the first cycle of the $F(X)$ dependence,
which defines the lowest value of $F^{\prime}(X)$ because it
determines the stability limit of the smooth sliding regime according
to Eq.~(\ref{Q16}).
As shown in Fig.~\ref{A08e},
higher  temperatures (Fig.~\ref{A08e}a) raise the minimum
value of $F^{\prime}(X)$,
which in turn extends the interval of the model parameters for which the
smooth sliding takes place.
At the same time, the higher is driving velocity (Fig.~\ref{A08e}b),
the smaller is the minimum of $F^{\prime}(X)$,
so that the narrower is the interval of model parameters where the
smooth sliding regime exists.
Thus, we come to a surprising conclusion that, at $T>0$,
reducing the pulling velocity may lead to a transition from stick-slip
to smooth sliding or creep motion, while one generally expects that
smooth sliding is reached when the velocity increases.
However a transition to smooth motion by reducing velocity
has also been observed experimentally~\cite{YGI1993}.


The behavior of $F_k$ on $T$ and $v$ described above is only
qualitative. A quantitative study to identify the temperature range in
which these effects would play a significant role in an actual
material would require a specific study. However it qualitatively agrees
with  tip-based experiments \cite{SO2003,SJHF2006}, which suggest
that, in these experiments, the tip/substrate contact
occurs through many atoms, i.e.\ it actually corresponds to multiple contacts.
Surprisingly, the dependences obtained within the ME approach,
perfectly agree with the experimental ones
for a model tribological
system~\cite{MIKOT2005,MIKSOTN2005,MN2007,NKKGMKB2007},
where the wearless kinetic friction of the driven lattice of
quantized magnetic vortexes in high-temperature cuprate superconductors
was studied
(e.g., compare inset of Fig.~\ref{A08c}b with Fig.~3 of Ref.~\cite{MIKOT2005}).

However our approach only claims to explain the general trends of the behavior
of tribological systems, because we neglected several important factors.
First, we neglected contact aging (see Sec.~\ref{aging}).
Second, we assumed that $R(x)=\delta (x)$, while at $T>0$ the
distribution $R(x)$ of stretchings when contact form again after a slip event
should correspond to the Boltzmann distribution around $x=0$
with a width depending on $T$~\cite{FKU2004,SW2009}.
Third, in a real system the contacts are heated due to sliding and even may
change their structure.
Forth, we ignored the elastic interaction between the contacts.
Finally, in macroscopic experiments the wearing of substrates in sliding contact
may mask the temperature and velocity dependences.
A more detailed investigation of these aspects deserves separate studies.

%------------------------------------------------------------------
\section{Aging of the contacts}
\label{aging}

In the approach described in Sec.\ \ref{analytics}, % above,
the function $F(X)$ was independent on the driving velocity
because we neglected two time-dependent effects: first a broken
contact needs a minimum delay $\tau_m$ to stick again, second we
ignored the aging of the contacts once they are formed, which
contradicts most of the experimental results
\cite{P0,BN2006,BC2006}.
Experiments as well as molecular dynamics simulations indicate that the static
friction
force grows with the lifetime $t_w$ of stationary contacts, i.e.\
the time interval during which they stay static prior to sliding.

\smallskip
The simplest way to obtain a velocity dependence of
the function $F(X)$ is to take into account the delay time $\tau_m$.
This is considered in Appendix~\ref{delay}.
In the most general case the delay time may not be the same for all contacts
but exhibits some distribution.
In Ref.~\cite{BT2009} we showed that the condition $\tau_m >0$
is the second necessary condition for the appearance of stick-slip.

\smallskip
Let us now consider the aging of contacts which requires a more involved
analysis because we must take into account a time evolution of the
distribution $P_c(x)$ of the contact breaking thresholds.
Let the newborn contacts be characterized by a distribution $P_{ci} (x)$
with the average value $\bar{x}_{si}$ and the dispersion $\sigma_{si}$.
If aging effects are taken into account,
then typically $\bar{x}_{s}$ grows with time, while $\sigma_{s}$
decreases, because the area of contact slowly grows under pressure.
As it does it faster
for the small contacts, which are less resilient, this tends to reduce
the dispersion in the properties of the contacts.
As a result, at $t \to \infty$ the distribution $P_c (x)$ approaches
a distribution $P_{cf} (x)$ with
$\bar{x}_{sf} > \bar{x}_{si}$ and
$\sigma_{sf} < \sigma_{si}$. The final distribution is a property of
the material. It is therefore a known quantity for the model once the
material has been specified.
If we assume that the evolution of $P_c (x)$ corresponds to a
stochastic process,
then it should be described by a Smoluchowsky equation
% $\partial P_c /\partial t = D \, \widehat{L}_f P_c$,
\begin{equation}
\frac{\partial P_c}{\partial t} = D \,  \widehat{L}_x P_c,
\;\;\; {\rm where} \;\;\;
\widehat{L}_x \equiv \frac{\partial }{\partial x} \left( B(x) +
\frac{\partial }{\partial x} \right),
\label{L}
\end{equation}
in which the ``diffusion'' parameter $D$ describes the rate of aging,
$B(x)=d\widetilde{U}(x)/dx$,
and the ``potential'' $\widetilde{U}(x)$ defines the final distribution,
$P_{cf} (x) \propto \exp \left[ -\widetilde{U}(x) \right]$;
therefore we can write
\begin{equation}
B(x) = -\frac{ dP_{cf} (x) /dx }{P_{cf} (x)}.
\label{B}
\end{equation}
Then, the equation
$\partial P_c /\partial t = D \, \widehat{L}_x P_c$
naturally leads to a growth of the average static threshold
% naturally leads to an exponential growth of the average static threshold
% $F_s (X) = \int_{-\infty}^{\infty} df \, f P_c(f;X)$
from $\bar{x}_{si}$ to $\bar{x}_{sf}$ with the time of stationary contact,
as widely assumed in earthquake-like models of friction
\cite{BN2006,P1995,BR2002}.
A characteristic timescale of aging may be estimated as
$\tau_{\rm aging} = (\bar{x}_{sf} - \bar{x}_{si})^2 /D$.

\begin{figure}[h] %[t] \bigskip
\includegraphics[width=9cm]{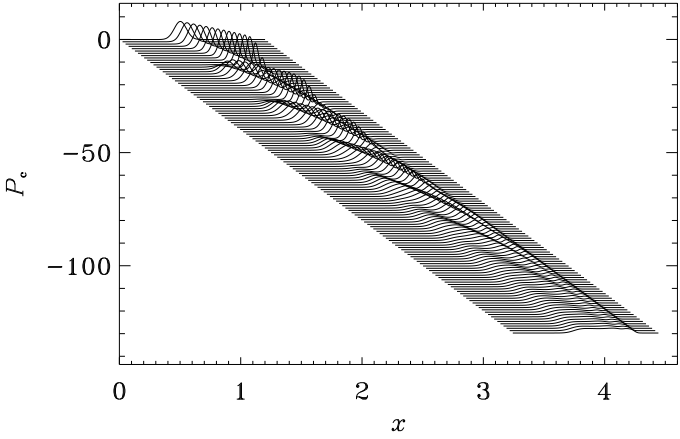}
\caption{\label{A04}
Evolution of $P_c (x)$ due to two concurrent processes:
the first process is the aging of asperities from the initial (fresh)
distribution
$P_{ci} (x)$ (Gaussian with $\bar{x}_{si}=0.5$ and $\sigma_{si}=0.05$) to
the final distribution
$P_{cf} (x)$ (Gaussian with $\bar{x}_{sf}=1$ and $\sigma_{sf}=0.02$),
and the second one is the breaking/re-binding of the contacts. $D_X
\equiv D/v =5 \times 10^{-4}$,
the initial distribution $Q_{\rm ini} (x)$ is
Gaussian with $\bar{x}_{\rm ini} =0.1$ and $\sigma_{\rm ini} =0.025$.
The curves are plotted with the increment $\Delta X = 0.05$.
%[use A04 = earth08e.pro]
}
\end{figure}
\begin{figure}[h] %[t] \bigskip
\includegraphics[width=8cm]{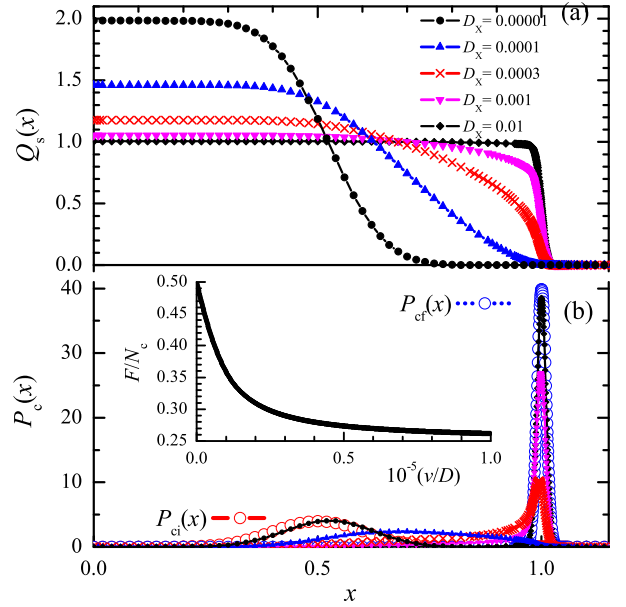}
\caption{\label{A05} (Color online):
(a)~The final distribution $Q_s (x)$ for five values
of the parameter $D_X =D/v$:
$10^{-5}$ (black circles),
$10^{-4}$ (blue up triangles),
$3 \times 10^{-4}$ (red crosses),
$10^{-3}$ (magenta down triangles), and
$10^{-2}$ (black diamonds).
(b)~The corresponding distributions $P_c (x)$.
Larger open circles show the initial distribution $P_{ci} (x)$ (red dashed) and
the final distribution $P_{cf} (x)$ (blue dotted).
% distributions in Eqs.~(\ref{B}, \ref{Q15})].
The inset displays the dependence of the kinetic friction force $F_k$
in the steady state on the driving velocity $v$ ($\langle k_i \rangle =1$).
$P_{ci} (x)$ is Gaussian with $\bar{x}_{si}=0.5$ and $\sigma_{si}=0.1$, and
$P_{cf} (x)$ is Gaussian with $\bar{x}_{sf}=1$ and $\sigma_{sf}=0.01$.
%[use A05]
}
\end{figure}

At the same time, the contacts continuously break and form again when the
substrate moves, as described by third and forth terms in Eq.~(\ref{Q5a}).
This introduces two extra contributions in the equation determining
$\partial P_c(x;X) / \partial X$ in addition to the pure aging effect described
by Eq.~(\ref{L}): a term $P(x;X) \, Q(x;X)$ takes into
account the contacts that break, while their reappearance with the threshold
distribution $P_{ci}(x)$ gives rise to the second extra term in the equation.
Combining the equation describing the evolution of the distribution
$Q(x;X)$ with the equation describing the aging of the contact
properties, and taking into account the
relation (\ref{PcPx2}),
we come to the system of three equations:
\begin{eqnarray}
&& \frac{ \partial Q(x;X) }{ \partial x}  +
\frac{ \partial Q(x;X) }{ \partial X} +
P(x;X) \, Q(x;X)
\nonumber \\
&& \;\;\;\;\;\;\;
= \delta(x) \int_{-\infty}^{\infty} dx^{\prime} \, P(x^{\prime};X) \, Q(x^{\prime};X),
\nonumber \\
&& \frac{\partial P_c (x;X)}{\partial X} - D_X \widehat{L}_x P_c (x;X)
+ P(x;X) \, Q(x;X)
\nonumber \\
&& \;\;\;\;\;\;\;
=  P_{ci}(x) \int_{-\infty}^{\infty} dx^{\prime} \, P(x^{\prime};X) \, Q(x^{\prime};X),
\nonumber \\
&& P_c (x;X) = % P(x;X) \, E_P (x;X) = % \nonumber \\ % && =
P(x;X) \, \exp \left[ - \int_{0}^{x} d\xi \, P(\xi;X) \right], % \;\;\; (f>0).
\label{Q15}
\end{eqnarray}
where $D_X = D/v$, and $v=dX(t)/dt$ is the driving velocity.
Thus, in the case of high driving ($v \to \infty$, $D_X \to 0$)
we recover the previous behavior with $P_c (x) = P_{ci} (x)$.
In the case of low driving ($v \to 0$, $D_X \to \infty$)
we again observe the same type of behavior but with $P_c (x) = P_{cf} (x)$,
and in the case of $0 < D_X < \infty$ we have an interplay of two processes,
the aging which tends to change $P_c (x)$ towards $P_{cf} (x)$,
and the breaking-reformation of the contacts which returns
$P_c (x)$ to $P_{ci} (x)$.
This is illustrated in Fig.~\ref{A04}, where we plot the evolution of $P_c (x)$
when $X$ continuously increases.
As a result, the friction force $F$ has to depend on the sliding velocity
as shown in inset of Fig.~\ref{A05}b.
Because typically $\bar{x}_{si} < \bar{x}_{sf}$,
the kinetic friction force $F_k$ generally decreases when $v$
grows, so that $dF_k (v)/dv <0$,
which may lead to an instability of the smooth sliding regime.

\smallskip
The steady-state solution of Eqs.~(\ref{Q15}) can be found analytically
in the low- and high-velocity cases.
In the $v \to 0$ limit,
\begin{eqnarray}
 P_c (x) \approx P_{cf} (x) + \frac{v}{D} \frac{P_{cf} (x)}{C[P_f]} \times
\nonumber \\
 \int_0^{x} d\xi \; P_{cf}^{-1} (\xi)
\int_0^{\xi} d\xi' \; \left[ P_{cf} (\xi') - P_{ci} (\xi') \right],
\nonumber
\end{eqnarray}
which leads to  $F_k (v) - F_k (0) \propto - v/D$,
while in the high-velocity case, $v \to \infty$, we have
\[
 P_c (x) \approx P_{ci} (x) + \frac{D}{v} \; C[P_i]
 \frac{d}{dx} \left[ B(x) + \frac{d}{dx}  \right] P_{ci} (x),
\]
which gives $F_k (v) - F_k (\infty) \propto D/v$. In these expressions
$C[P]$ designates the constant $C$ defined by Eq.~(\ref{Q11c}), which
takes different values depending on the expression of $P(x)$.
Both limiting cases may be combined within one fitting formula,
\begin{equation}
F_k (v) \approx F_k (\infty) + \frac{F_k (0) - F_k(\infty)}{1+v/v_{\rm aging}},
\label{fitFv}
\end{equation}
where $v_{\rm aging}$ defines the velocity below which aging effects
become essential. 
For example, for the parameters used in Fig.~{\ref{A05}} we found that
$v_{\rm aging} \approx 6.5 \times 10^3 D$.

\smallskip

\begin{figure} % [t] %[ht] \bigskip
\includegraphics[width=8cm]{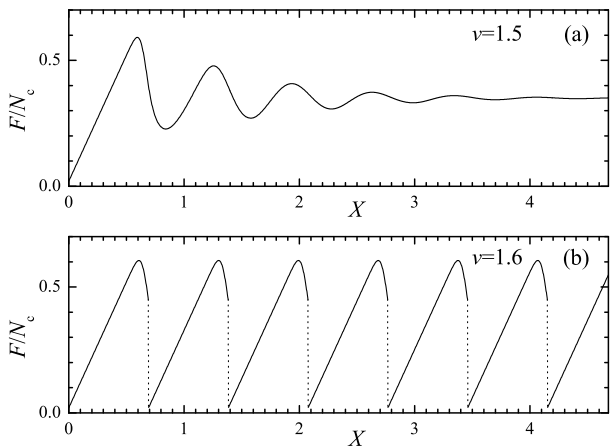}
\caption{\label{A06}
The dependences $F(X)$ for two driving velocities
$v=1.5$ (a) and $v=1.6$ (b).
The initial (fresh) distribution of thresholds $P_{ci} (x)$
is Gaussian with $\bar{x}_{si}=0.5$ and $\sigma_{si}=0.05$,
the final distribution $P_{cf} (x)$
is Gaussian with $\bar{x}_{sf}=1$ and $\sigma_{sf}=0.02$.
The initial distribution for contacts $Q_{\rm ini} (x)$ is
Gaussian with $\bar{x}_{\rm ini} =0.02$ and $\sigma_{\rm ini} =0.01$
(for calculation reasons, the same is used for the ``refreshed''
distribution after the break
at the point $F^{\prime}(X)=-K$).
The parameters are $\langle k_i \rangle =1$, $K=3.5 N_c \langle k_i
\rangle$, and
the rate of aging is determined by the coefficient $D_X =D/v$ with $D
=5 \times 10^{-4}$.
%[use A06]
}
\end{figure}
When the sliding substrates are not rigid, the effect of contacts aging may
lead to a transition from stick-slip to smooth sliding with the increase of
the sliding velocity.
When the driving velocity increases and reaches
a critical value $v=v_c$ the transition is abrupt (first-order)
as demonstrated in Fig.~\ref{A06}.
Moreover, the system exhibits hysteresis:
 starting from the smooth-sliding regime, if the velocity decreases
we expect a smooth sliding to survive
down to velocities much lower than $ v_c$.

In a general case, the parameter $D$ in Eq.~(\ref{Q15}) and, therefore
the rate of contacts aging, depends on the system temperature $T$.
Typically, the aging rate should increase with $T$ according to Arrhenius law,
$D \propto \exp \left( -\varepsilon /k_B T \right)$, where
%$k_B$ is Boltzmann's constant and
$\varepsilon$ is an activation energy for this process.
Besides, aging mechanisms may be much more involved than
the simplest one described by the Smoluchowsky equation~(\ref{L})
and may even consist of two stages~\cite{BC2006}, the ``geometrical''
growth of contacts
(described, e.g., by the Lifshitz-Sl\"ozov theory, see Sec.~\ref{Lifshitz})
and then a structural reordering of asperities.

%------------------------------------------------------------------
\section{Origin of threshold distribution}
\label{disc}

As discussed in the introduction the formalism using the master
equation to describe friction allows us to separate the calculation of
the friction force, assuming that the statistical properties of the
contacts are known, from the analysis of
the properties of the contact themselves. In
the previous sections we focused on the calculation of the friction
force, which is our main goal. However it is interesting to examine
the properties of the contacts because it allows us to draw some
general conclusions on friction, particularly on the possible
existence of stick-slip. Within our approach the properties of the
contacts are entirely described within the distribution $P_c (x)$ of
the contact thresholds, which may be time dependent as discussed in
Sec.~\ref{aging} on aging.
Let us now discuss possible physical origins of this distribution.

To establish the master equation, it is convenient to express the
properties of the contacts in term of their stretching $x$, hence
$P_c(x)$, but at the microscopic scale contacts are generally
characterized by their breaking force. The distribution $P_c(x)$
can be related to the distribution $\widetilde{P}_c (f_s)$
of the static friction force thresholds of the contacts.
If a given contact has an area $A$, then it is characterized by the
static friction threshold $f_s \propto A$ and
the elastic constant $k \sim \rho \, c^2 \sqrt{A}$,
where $\rho$ is the mass density and $c$ the sound velocity
(assuming that the linear size of the contact and its height
are of the same order of magnitude~\cite{P1995}).
The displacement threshold for the given contact is $x_s = f_s /k$,
so that
\begin{equation}
f_s \propto x_s^2,
\label{fsvsxs}
\end{equation}
or $df_s /dx_s \propto x_s$.
Then, using the relationship $P_c (x_s) \, dx_s = \widetilde{P}_c (f_s) \, df_s$,
we obtain
\begin{equation}
P_c (x_s) \propto x_s \widetilde{P}_c [f_s (x_s)].
\label{PxPf}
\end{equation}

%------------------------------------------------------------------
\subsection{Rough surfaces: the GW model}
\label{rough}

Let us consider the contact of two nominally flat surfaces, for which
the appararent contact area is large so that individual contacts are
dispersed and the forces acting through neighboring spots do not
influence each other \cite{GW1966}. Their study can be cast into the
problem of the contact of a rigid plane and a
composite surface whose topography is the sum of both
topographies, with an appropriate renormalization of some parameters
such as the Young modulus, so that
the contact problem reduces to the statistics of independent
asperities~\cite{BC2006}.
Let the rough surface be characterized by hills of heights $\{ h_i \}$
distributed with a probability $P_h (h)$.
Following the Greenwood and Williamson (GW) model of the interface
\cite{GW1966},
let us assume that all hills have a spherical shape of the same
radius of curvature $r$.
When this surface is pressed with another rigid flat surface,
which takes a position at the level $h_0$,
then the hills of heights $h>h_0$ will form contacts, or asperities.
If the contacts are elastic (the so called Hertz contacts \cite{J1985}),
then the contact of height $h$ has the compression
$(h-h_0)$, its area is $\pi r (h-h_0)$, and it bears the normal force
$f_{l} (h) \approx (4\pi /3) E^* r^{1/2} (h-h_0)^{3/2}$,
where $E^*$ is the effective Young modulus
(if both contacting substrates have the same Young modulus $E$, then
$E^*=E/2(1-\nu^2)$, where $\nu$ is the Poisson modulus \cite{J1985}).
It is reasonable to assume that the shear static threshold for the given contact
is proportional to the load force, $f_{s} (h) \approx \mu f_{l} (h)$, or
\begin{equation}
f_{s} = (4\pi /3) \mu E^* r^{1/2} (h-h_0)^{3/2},
\label{GW1}
\end{equation}
where $\mu \alt 1$ is a constant.
Then the distribution of static thresholds $\widetilde{P}_c (f_s)$ can
be coupled with
the distribution of asperities heights $P_h (h)$
with relation $\widetilde{P}_c (f_s) \, df_s \propto P_h
(h) \, dh$, or
$\widetilde{P}_c (f_s) \propto (dh/df_s) P_h \left[ h(f_s) \right]$, where
$dh/df_s \propto f_s^{-1/3}$ according to Eq.~(\ref{GW1}).

\begin{figure}[h] %[t] \bigskip
\includegraphics[clip,width=8cm]{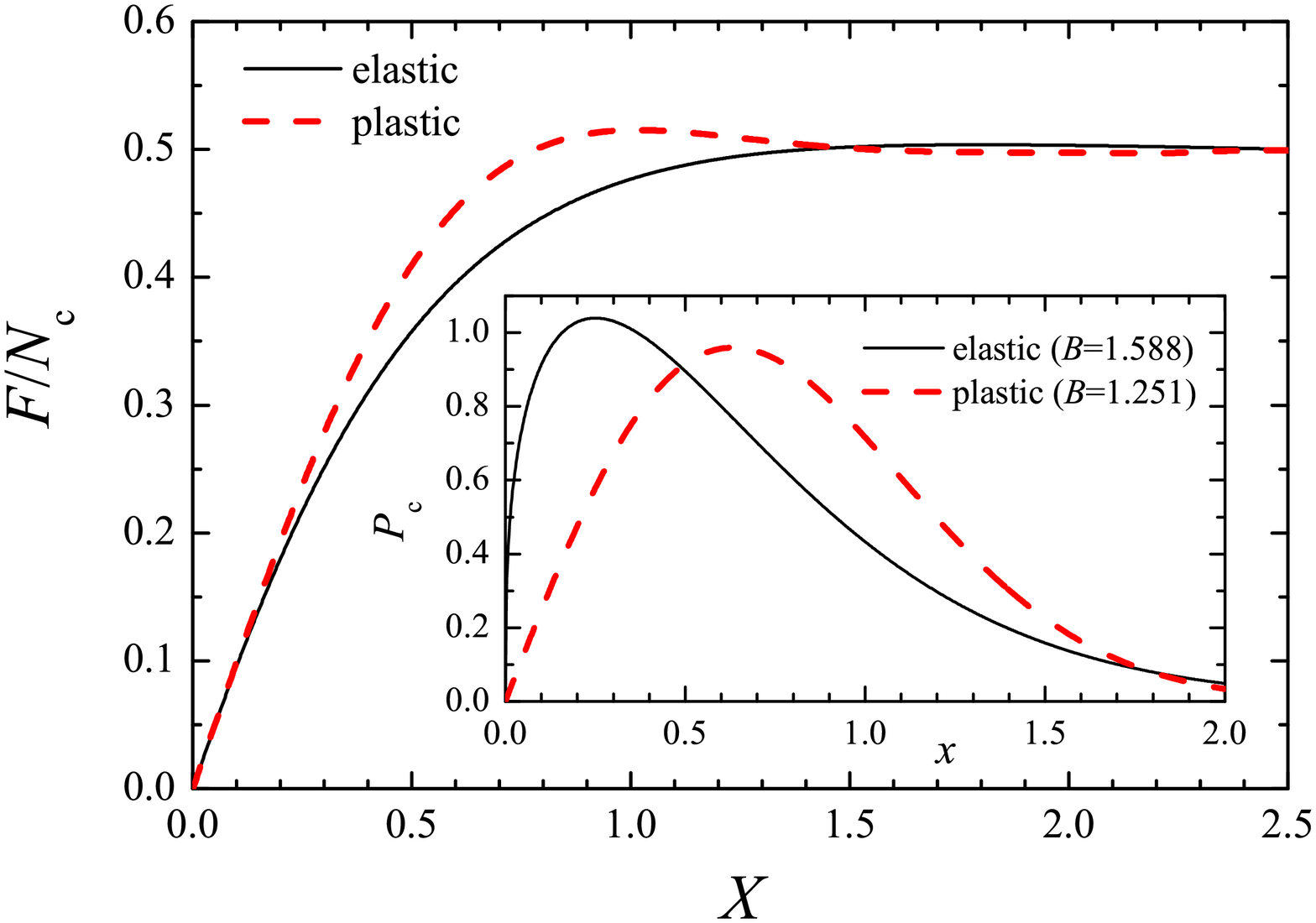} % clip
\caption{\label{A10a} (Color online):
Dependence of the friction force $F$ on $X$ for elastic
($B^{\prime}=1.588$, solid curve)
and plastic ($B^{\prime \prime}=1.251$, broken curve) contacts
with exponential distribution of heights.
The initial distribution $Q_{\rm ini} (x)$ is
Gaussian with $\bar{x}_{\rm ini} =0$ and $\sigma_{\rm ini} =0.025$.
The inset shows the corresponding distributions $P_c (x)$,
Eqs.~(\ref{GW3}) and~(\ref{GW4}).
%[use Rotate/idl/Pexperimental/Pc(x)-1.opj = A10a]
}
\end{figure}

\begin{figure}[h] %[t] \bigskip
\includegraphics[clip,width=8cm]{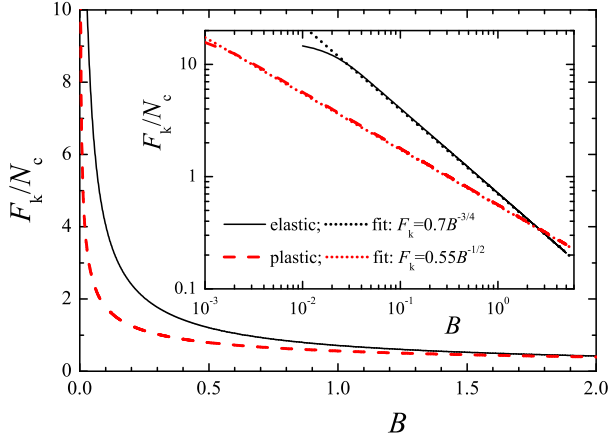} % clip
\caption{\label{A10b} (Color online):
Dependence of the kinetic friction $F_k$ in the smooth sliding state
on the parameter $B$ for the elastic (solid curve) and plastic (broken
curve) contacts.
Inset: the same in log-log scale; dotted lines show power-law fits.
%[use Rotate/idl/Pexperimental/Pc(x)-1.opj = A10b]
}
\end{figure}

\smallskip
For a strong load, when the local stress exceeds the yield threshold $Y$,
the contacts begin to deform plastically.
When all contacts are plastic, the local pressure on contacts is
$p_{\rm load} = H$, where $H$ is the hardness
[$H \approx 3Y$ for the spherical geometry of asperities;
for metals $H \approx (10^{-3} - 10^{-2})E \sim 10^9$~Pa].
Then the normal force at the contact is $f_{l} (h) \approx \pi r (h-h_0) H$.
Assuming again that $f_{s} (h) \approx \mu f_{l} (h)$, we obtain
\begin{equation}
f_{s} = \pi r (h-h_0) \mu H,
\label{GW2}
\end{equation}
so that $\widetilde{P}_c (f_s) \propto P_h \left[ h(f_s) \right]$ now.

\smallskip
For example, if we consider the exponential distribution of heights
introduced by Greenwood and Williamson \cite{GW1966} to get analytical results
$P_h (h) = \bar{h}^{-1} \exp (-h/\bar{h}) \, \Theta (h)$,
where $\bar{h}$ is the average height,
then with Eq.~(\ref{PxPf}) for the elastic contacts we
obtain ($f,x>0$)
\begin{eqnarray}
\widetilde{P}_c (f) \propto f^{-1/3} \exp (-B_1 f^{2/3}),
\;\;\; {\rm or}
\nonumber \\
P_c (x) \propto x^{{1}/{3}} \exp (-B^{\prime} x^{4/3}),
\label{GW3}
\end{eqnarray}
where
$B_1 = \left[ (4\pi /3)^{2/3} \bar{h} r^{1/3} (\mu E^*)^{2/3} \right]^{-1}$,
while for the plastic contacts
\begin{equation}
\widetilde{P}_c (f) \propto \exp (-B_2 f),
\;\;\; {\rm or} \;\;\;
P_c (x) \propto x \exp (-B^{\prime \prime} x^{2}),
\label{GW4}
\end{equation}
where
$B_2 = (\pi \bar{h} r \mu H)^{-1}$. $B^{\prime}$ and
$B^{\prime\prime}$ are numerical constants depending on the material.

The distributions $P_c (x)$ for elastic and plastic contacts,
Eqs.~(\ref{GW3}) and~(\ref{GW4}), are presented in Fig.~\ref{A10a} (inset).
We see that the force $F(X)$ increases (almost) monotonically with $X$,
approaching the kinetic value $F_k$.
Thus, in the case of a contact of rough surfaces, a relatively high
concentration of low-threshold contacts prevents the stick-slip motion
even for a very soft tribosystem (i.e., even in the case of
a very low elastic constant of the driving spring $K$).
The kinetic friction force $F_k$ depends on the parameter $B$
according to the power law (see Fig.~\ref{A10b}),
$F_k \propto B^{-3/4} \propto \bar{h}^{3/4}$
for the elastic contacts, and
$F_k \propto B^{-1/2} \propto \bar{h}^{1/2}$
for the plastic contacts.

Similarly one can find $\widetilde{P}_c (f)$ for a more realistic
Gaussian distribution of heights of the asperities \cite{GW1966}.
In this case, the peak in the $P_c (x)$ dependence
becomes narrower and higher,
and the stick-slip regime may exist.
Note that typical values for rough surfaces are of the order
$r \sim 10 - 100$~$\mu$m, $\bar{h} \sim 1$~$\mu$m, so that the average
size of the contact is
$\bar{a} \approx \left( r \bar{h} \right)^{1/2} \sim 3 - 10$~$\mu$m
\cite{BC2006}.

Whether the contacts are in the plastic or elastic regime, depends on
the dimensionless parameter $\psi = (E/Y) (\bar{h}/r)^{1/2}$: $\psi
\gg 1$, which is typical for metals, corresponds to the plastic regime
while $\psi < 1$ leads to an elastic regime, as, for instance, for rubber
friction.  Polymeric glasses belong to an intermediate situation,
$\psi \agt 1$, where only a fraction of contacts is plastic
\cite{BC2006}.

%------------------------------------------------------------------
\subsection{Rough surfaces: Persson's model}
\label{roughPersson}

The GW model used above completely ignores the elastic interaction
between the contacts thus overestimating the role of low-threshold contacts.
Recently Persson~\cite{P2001a} developed a contact mechanics theory
which (indirectly) includes elastic interactions and leads to the correct
low-threshold limit $\widetilde{P}_c (f \to 0) =0$.
Although Persson's theory is only rigorous for the case of complete contact
(e.g., contact of rubber surfaces at high load),
it leads to a very good agreement with experiments and simulations
even for a contact of stiff surfaces and low loads~\cite{YP2008}.
Not going into details, note that the distribution
of normal stresses $\sigma_n$ at the interface
can approximately be described by an expression ($\sigma_n >0$)
\begin{equation}
\label{PcPersson}
P_{n} ({\sigma_n}) \propto
\exp \left[ - \left( \frac{\sigma_n - \bar{\sigma}_n}{\Delta \sigma_n}
  \right)^2 \right] -
\exp \left[ - \left( \frac{\sigma_n + \bar{\sigma}_n}{\Delta \sigma_n}
  \right)^2 \right],
\end{equation}
where $\bar{\sigma}_n$ is the nominal squeezing pressure,
the distribution width is given by $\Delta \sigma_n = E^* {\cal R}^{1/2}$,
and the parameter ${\cal R}$ is determined by the roughness of the
contacting surfaces,
\begin{equation}
\label{PcPersson2}
{\cal R} = \frac{1}{4\pi} \int dq \; q^3 \int d^2 x \; \langle h({\mathbf x})
h({\mathbf 0}) \rangle e^{-i {\mathbf q} {\mathbf x}}.
\end{equation}
Assuming that a local shear threshold is directly proportional to the
local normal stress,
$f_s \propto \sigma_n$,
and again using Eq.~(\ref{PxPf}), we finally obtain
\begin{eqnarray}
\label{PcPersson3}
P_c (x) \propto x \Bigg\{ &&
\exp \left[ - \left( \frac{x^2 - \bar{x}^2}{\sigma^2} \right)^2
  \right] \nonumber \\&&-
\exp \left[ - \left( \frac{x^2 + \bar{x}^2}{\sigma^2} \right)^2 \right]
\Bigg\},
\end{eqnarray}
where $\sigma \propto (E^*)^{1/2} {\cal R}^{1/4}$.
The distribution (\ref{PcPersson3}) is characterized by a low concentration
of small shear thresholds,
$P_c (x) \propto x^3$ at $x \to 0$,
and a fast decaying tail,
$P_c (x) \propto \exp [-(x/\sigma)^4]$ at $x \to \infty$,
i.e., the peaked structure of the distribution (\ref{PcPersson3})
is much more pronounced than in the GW model, Eqs.~(\ref{GW3}) and (\ref{GW4}).

\begin{figure}[h] %[t] \bigskip
\includegraphics[clip,width=8cm]{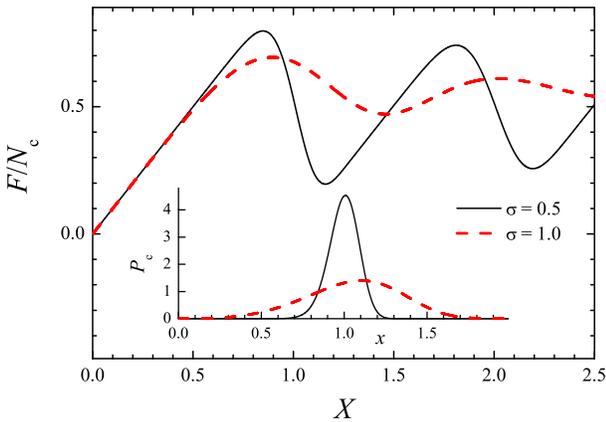} % clip
\caption{\label{figPcPersson} (Color online):
Dependence of the friction force $F$ on $X$
for the Persson model with $\bar{x}=1$,
$\sigma = 0.5$ (solid) and $\sigma = 1$ (broken curve).
The initial distribution $Q_{\rm ini} (x)$ is
Gaussian with $\bar{x}_{\rm ini} =0$ and $\sigma_{\rm ini} =0.025$.
The inset shows the corresponding distributions $P_c (x)$,
Eq.~(\ref{PcPersson3}).
%[use figPcPersson.opj]
}
\end{figure}

The numerical results are presented in Fig.~\ref{figPcPersson}
for two values of the roughness parameter. In contrast to the results
of the GW model, they show a non-monotonous behavior of $F(X)$ which
is compatible with a stick-slip motion.

%------------------------------------------------------------------
\subsection{Flat surfaces: dry friction}
\label{dry}

Let us now consider the dry contact of two flat surfaces.
If both surfaces have an ideal crystalline structure,
we get the singular case discussed in Appendix~\ref{singular}.
However such a situation is exceptional.
A real surface always consists of domains, which are characterized by
different crystalline orientation or even different structures. 

MD simulations show a large variation of the static friction
with the relative orientation of the two bare substrates
\cite{HS1993,RS1996}, although
surface irregularities as well as fluctuations of atomic positions
at nonzero temperatures may mask this dependence~\cite{QCCG2002}.
A variation of the friction with the misfit angle was also
observed experimentally, for instance in
the FFM experiment made by Dienwiebel
\textit{et al.} \cite{DVPFHZ2004}.

In order to estimate a possible shape of the function $P_c (x)$
resulting from the domain structure of substrates,
let us consider a rigid island (domain)
of size $N_a$ with a triangular lattice
placed onto a 2D periodic potential with square symmetry $V(x,y)=
\sin x + \sin y$  created by the bottom substrate.
The atomic coordinates of domain atoms are
$\tilde{x}_i = X + (i+j/2)a$ and $\tilde{y}_j = Y + jh$,
where $h=a \sqrt{3}/2$,
$i=-j/2,\ldots,-j/2+n_i-1$,
$j=0,\ldots,n_j-1$,
$n_i n_j =N_a$
($n_i a \approx n_j h$),
and $X,Y$ are the center of mass coordinates.
If the island is rotated by an angle $\phi$, then the atomic
coordinates change to 
$x_i = \tilde{x}_i \cos \phi - \tilde{y}_i \sin \phi$ and
$y_i = \tilde{x}_i \sin \phi + \tilde{y}_i \cos \phi$.
For a fixed misfit angle $\phi$, the total potential energy of the domain is
$U(X,Y)=\sum_{ij} V(x_i,y_j)$.
Extrema (minima and saddle points) of the function $U(X,Y)$ are
defined by 
$\partial U / \partial X = \partial U / \partial Y =0$;
then the activation energy for domain motion
is given by $\varepsilon_a = U_{\rm saddle} - U_{\rm min}$.
Assuming that $f_s \sim \varepsilon_a /a_s \propto \varepsilon_a$,
we can estimate the threshold force $f_s$
as a function of the misfit angle $\phi$ and the domain size $N_a$.
This protocol was realized by Manini and Braun~\cite{BM2010b}.
The calculation of a histogram of the function $\varepsilon_a (\phi)$
leads to the distribution $\widetilde{P}_c (f_s)$, assuming that
all domains have the same size $N_a$ and that all angles are equally present.
Next, one may average over different domain sizes $N_a$, e.g.,
with a weight function $w(N_a) = e^{-N_a /\bar{N}_a}$,
where $\bar{N}_a$ is the average domain size.
The distribution $\widetilde{P}_c (f)$ obtained in this way may be
crudely approximated by the function
$\widetilde{P}_c (f) \propto \exp (- f/\bar{f})$.
Assuming that Eq.~(\ref{PxPf}) holds, we come to the distribution
\begin{equation}
P_c (x) \propto x \exp (- x^2 / \sigma^2) \; ,
\label{Pcrotateh}
\end{equation}
which is similar to that of Fig.~\ref{A10a} characteristic for rough surfaces.
Therefore in this system we again expect smooth sliding without stick-slip
due to a large concentration of configurations with small barriers.

However, in the estimation we supposed
that all angles are present with equal probability,
while some angles could have preference due to their lower potential energy.
Calculation shows that this would suppress the low-barrier thresholds 
so that the distribution $P_c (x)$ would become more peaked,
and the stick-slip could exist.

%------------------------------------------------------------------
\subsection{Lubricated surfaces: MD simulation}
\label{LSsimulation}

The dry-friction considered above is exceptional.
In a real system, there is almost always a lubricant between the
sliding surfaces
(called ``the third bodies'' by tribologists) which is
either a specially chosen lubricant film, grease (oil),  dust,
wear debris produced by sliding, or water or/and a thin layer of hydrocarbons
adsorbed from air.

There is a large number of experimental and simulation studies of
lubricated friction
(e.g., see Refs.~\cite{P0,BN2006} and references therein).
A thin confined film (less than about six molecular diameters of thickness)
typically solidifies thus producing nonzero static friction.
The value of $f_s$ strongly depends on the film thickness and its structure,
and moreover it may change with the time of stationary contact.
In particular,
Jabbarzadeh \textit{et al.\/} \cite{JHT2006} have done the
MD simulation of a thin lubrication film of dodecane.
The transition from bulk liquid to high viscosity state when thickness
decreases
occurs at the thickness of six lubricant layers, and it appears
due to formation of isolated crystalline bridges between the mica
surfaces (across the film).
As the thickness decreases further, these bridges increase in number and
organize themselves into a mosaic structure with a long range
orientational  order. However in the previous MD study
of the six-layer dodecane film between the mica substrates
Jabbarzadeh \textit{et al.\/} \cite{JHT2005}
found also a ``layer-over-layer'' sliding regime with a very low friction.
Such a configuration is found to be thermodynamically stable contrary to the
metastable ``bridge'' configuration mentioned above.
Thus, if the film is not ideally homogeneous along the interface
but consists of domains of different structures and may be
different thicknesses,
it should be characterized by a distribution of static thresholds.

\smallskip
Moreover the top and bottom surfaces may be misoriented as well, as
was mentioned above.
In particular, He and Robbins studied the dependence of the static
\cite{HR2001s}
and kinetic \cite{HR2001k} friction on the rotation angle of the substrates
for a lubricated system. It was observed that
the static friction exhibits a peak at the commensurate angle ($\phi=0$)
and then is approximately constant,
the peak/plateau ratio being about 7 (for the monolayer lubricant film,
for which the variation is the strongest).

A detailed study of the dependence of the static friction on the
rotation angle $\phi$
for lubricant films of thickness from one to five layers
has been done by Braun and Manini~\cite{BM2010a}.
The value of $f_s$ varies with $\phi$ by more than one order in magnitude.
When averaged over $\phi$ (assuming that all angles are present with
the same probability)
and also over film thickness with some weighting factor,
the distribution of static thresholds may approximately be described
by the function
$\widetilde{P}_c (f) \propto f \exp (- f/\bar{f})$.
Assuming again that Eq.~(\ref{PxPf}) holds, we come to the distribution
\begin{equation}
P_c (x) \propto x^3 \exp (- x^2 / \sigma^2) \; ,
\label{Pcrotate}
\end{equation}
which would allow stick-slip.

%------------------------------------------------------------------
\subsection{Lubricated surfaces: Lifshitz-Sl\"ozov coalescence}
\label{Lifshitz}

\begin{figure}[h] %[t] \bigskip
\includegraphics[clip,width=8cm]{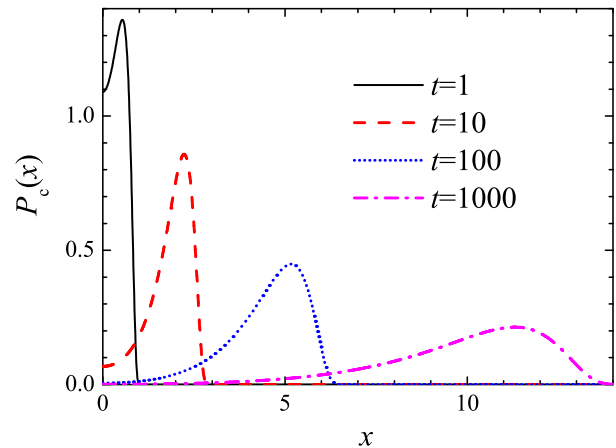} % clip
\caption{\label{A12} (Color online):
Evolution of the distribution $P_c (x)$ with the time of stationary contact
for the Lifshitz-Sl\"ozov coalescence mechanism
($\alpha=1$ and $B=1$).
%[use A12]
}
\end{figure}
In the conventional melting-freezing mechanism of friction,
the lubricant is melted during slip, and solidifies when the motion stops.
The solidification process can be described
by the Lifshitz-Sl\"ozov theory (see \cite{LS1958}, Sec.~100).
At the beginning, grains of solid phase emerge within the liquid lubricant film.
Then the grains grow in size
according to the law $\bar{r} \propto t^{1/3}$,
where $\bar{r}$ is the average grain radius.
The distribution of grains sizes is described by the following function:
the number of grains with the radius from $r$ to $r+\Delta r$
is equal to $P_{\rm LS} (r/\bar{r}) \, \Delta r/\bar{r}$, where
the function $P_{\rm LS}(u)$ is zero for $u \geq 3/2$
(so that the maximum size of the grains is $1.5 \bar{r}$),
while lower sizes are distributed as
\begin{equation}
P_{\rm LS}(u) \propto % = \left( \frac{3^4 e}{2^{5/3}} \right)
\frac
{u^2 \exp \left[ -1 / (1-2u/3) \right]}
{(u+3)^{7/3} \left( \frac{3}{2} -u \right)^{11/3}} \; . % ;
\label{LS1}
\end{equation}
% the function $P_{\rm LS}(u)$ is normalized by
% $\int_0^{\infty} du \, P_{\rm LS}(u)=1$.
Due to coalescence of grains,
the total number of grains decreases with time as $N(t) \propto t^{-1}$.
When the size of a grain exceeds the film thickness $d$ (that will occur
when $\bar{r} (t) \geq d/3$ after the delay time $\tau_m \propto d^3$),
such a grain will pin the surfaces.
Using the relationship $\widetilde{P}_c (f) df \propto P_{\rm LS}
(r/\bar{r}) \, dr/\bar{r}$,
we obtain that
$\widetilde{P}_c (f) \propto (dr/df) P_{\rm LS} (r/\bar{r}) /\bar{r}$.
Because one grain gives the static threshold proportional to the
contacting area,
$f_s \propto \pi (r^2 - d^2/4)$ for $r > d/2$, we have $dr/df \propto f^{-1/2}$.
Combining all things together, we obtain that
\begin{equation}
P_c (x)=P_{\rm LS}(u),
\;\;\;
u = \rho^{-1} \left( 1+B x^2 \right)^{1/2},
\label{LS2}
\end{equation}
where
$\rho (t) = 2 \bar{r}(t)/d = \alpha t^{1/3}$
(the pinning begins when $\rho > 2/3$),
and $B$ is determined by the system parameters
(elasticity of the contacts, proportionality between the threshold $f_s$
and the area of the contacting grain, and the thickness of the lubricant film).
The distribution (\ref{LS2}) is shown in Fig.~\ref{A12}.
Its shape suggests that
it should lead to the conventional tribological behavior:
the stick-slip motion at low driving velocity and smooth sliding at
high velocities.

Thus, the described mechanism leads to a non-singular distribution of
static thresholds
and the typical stick-slip to smooth sliding behavior even
for the case of ideal flat surfaces such as, e.g.,
the mica surfaces used in SFA experiments.
Note that it incorporates both the delay time discussed in Appendix~\ref{delay}
and the aging of contacts considered in Sec.~\ref{aging};
the latter, however, is a more complicated process than follows from
the simplest Smoluchowsky theory.

\smallskip
In a general case, we also have to take into account that the
lubricant film may consist
of grains (domains) of different orientations or even different structures.
Indeed, the proportionality coefficient in the relation
$f_s \propto \pi (r^2 - d^2/4)$
used above, should depend on the misfit angle between the lubricant
domain and the substrate, so that the distribution $P_{\rm LS} (u)$
introduced above, additionally should depend on the misfit angle $\phi$,
$P_{\rm LS} (u; \phi)$.
Thus, if there are grains with different orientations,
distributed according to a function $R_{\phi} (\phi)$,
then
\begin{equation}
P_c (x) = \int d\phi \, R_{\phi}(\phi) P_{\rm LS} (u; \phi).
\end{equation}
For example, if there are only two energetically equal possible orientations
with $\phi =0$ and $\phi =\pi/4$,
then $R_{\phi} (\phi) = {1\over2} \delta(\phi) + {1\over2} \delta(\phi -\pi/4)$.

%==================================================================
\section{Conclusion}
\label{conclusion}

We introduced and analyzed in detail the earthquake-like model
with a distribution of static thresholds.
Reducing the problem to a master equation,
we were able to find the exact solution numerically and even
analytically in some important cases.
Although, by its design, the ME approach should
    exactly correspond to the EQ model when the properties of the
    contacts are the same, it is difficult to give a formal proof of
    the equivalence of the two methods because the EQ model does not
    have a known analytical solution, except in very specific cases.
The smooth sliding, as well as the transition from stick-slip to smooth sliding,
emerge due to a finite width of the distribution $P_c (x)$
and have (almost) nothing in common with
the melting-freezing or inertia microscopic mechanisms of stick-slip.
We observed that, for stick slip to occur, the distribution of breaking
thresholds $P_c(x)$ must have a low enough weight for small thresholds,
for example, $P_c(x) \propto x^3$ if $ x \to 0$ as follows from
Eqs.~(\ref{PcPersson3}) and~(\ref{Pcrotate}).

The complex problem of behavior of a tribological system
is split into two independent subproblems.
In the present paper we studied the first one:
dynamics of the friction contact, if the distribution of static
thresholds $P_c (x)$ is known.
The second separate problem is to find the distribution
$\widetilde{P}_c (f)$ for a given system. Although we did not focus on
this problem we examined the various possible situations of
microscopic contacts to estimate $P_c(x)$ for those situations in
order to examine how they fit in the framework of the master equation
approach.
For a contact of two hard rough surfaces
(the multi-contact interface, or MCI in notations used by Baumberger
and Caroli~\cite{BC2006})
this problem reduces to the statistics of asperities.
The contact aging, i.e.\ the increase of static thresholds with
the time of stationary contact, is due to two processes:
the first (and more important) process is due to geometrical aging,
or the increase of contact area at the asperity,
and the second, due to structural aging, or restructuring of the contact.
The value of the parameter $D$ in Eq.~(\ref{Q15}) may be estimated
from experiments:
it was found that in most cases the average static threshold
$\mu_s = F_s /F_{\rm load}$ grows logarithmically with the waiting time $t_w$,
$d \mu_s /d \, (\ln t_w) \approx 10^{-2}$~\cite{BC2006}.

For flat surfaces such as, e.g., the mica surfaces used in the SFA experiments,
the surface consists, most probably, of domains of different orientations
(as for poly-crystals, and even for mono-crystals,
where simply the sizes of domains are much larger).
However, even if both sliding surfaces have an ideal crystalline structure
on the macroscopic scale, anyway the lubricant film should consist of domains
of different orientations or different metastable structures
separated by dislocations (domain walls).
The parameters of these domains can be studied by standard methods of
molecular dynamics,
while their evolution with time
(the structural aging, which should lead to the increase of the static
thresholds)
may be described, e.g., by the Ginzburg-Landau theory.

It is interesting to note that the EQ-ME viewpoint discussed here
appeared in biological physics more than fifty year ago
to describe skeletal muscles ~\cite{HS1971}.
The Lacker-Peskin model~\cite{P1975a,P1975b,LP1986,W1984}
describes the binding-unbinding of the myosin molecules on actin
filaments in terms of an equation which is very similar to our master
equation.

\medskip
Finally, let us mention questions which were not considered in the
present work.
First, throughout the paper we ignored inertia effects.
They would imply that the distribution $Q(x)$ should be extended
to negative $x$ values because some contacts could pick
up enough kinetic energy during sliding
to overshoot. Returning to the stick (pinned) state
the spring force acting on such a contact would be negative \cite{P1995}.
% Negligible (see Persson or Caroli)?
Moreover in the case of $dF_k (v)/dv <0$
which appears due to contacts aging,
the regimes of stick-slip and smooth sliding may be separated
by a regime of irregular (chaotic) motion due to inertia effects.
Second, we completely ignored the role of the interaction between
the contacts.
The effects of interaction should change (increase) the critical velocity
of the transition from stick-slip to smooth sliding.
Interaction may be incorporated indirectly in a mean-field fashion
by a renormalization of the distributions $P_c (x)$ and $R(x)$.
However, a concerted, or synchronized breaking (triggering) of
contacts may be studied
numerically only within the earthquake-like model (e.g., see
Ref.~\cite{BR2002}).
It cannot be included rigorously in the master equation approach,
unless it is coupled to a deformation field which deeply modifies the
approach.
Also, we considered the contacting planes of the sliding blocks
as rigid planes, i.e., we ignored a possible variation (or fluctuation) of $F$
in the $(x,y)$ plane. Moreover, the latter should be coupled with
the nonuniform elastic deformation of the block
(for a preliminary work in this direction see Ref.~\cite{BBU2009}).
Finally, we ignored a possible heating of the contacts due to sliding.
All these question deserve separate studies.
In conclusion, note that the approach developed in the present paper,
is to be applied to meso- or macroscopic systems only
and cannot be used to explain the AFM/FFM tip-based experiments,
except if there is a multi-contact through a sufficient number
of tip atoms.

%==================================================================
\acknowledgments
We wish to express our gratitude to N.\ Slavnov and T.\ Dauxois
for helpful discussions. We thank an anonymous referee for helpful
comments on the manuscript.
This work was supported by CNRS-Ukraine PICS grant No.~5421.

%==================================================================
\appendix
\section{Shear elastic constant}
\label{shearkc}

To estimate possible values of the shear constant $k_i$,
let us assume that the contacts have the shape of a cylinder of radius
$r_c$ and length $h$ (i.e., $h$ is the thickness of the interface).
Also, we suppose that one end of such a contact column is fixed,
while a force $f_i$ is applied to the free end.
This force will lead to the displacement $x_i = f_i/k_i$ of the end,
where $k_i = 3E_c I/h^3$, $E_c$ is the Young modulus of the material
of the contact
and $I = \pi r_c^4 /4$ is the moment of inertia of the cylinder
(see Ref.~\cite{LL1986}). 
% , Sec.~II.17, Eq.~(17.11), p.~97 and Sec.~II.20, exercise~3, p.~116
This leads to
\begin{equation}
\label{eq15}
k_i = (3\pi /4)(E_c r_c)(r_c /h)^3 \,.
\end{equation}
As the Young modulus is coupled with
the transverse sound velocity of the material
by the relationship % ~\cite{LL1986}
$E_c = 2 \rho (1+\sigma) c^2$,
where $\rho$ is the mass density and $\sigma$ is the Poisson ratio,
we obtain
\begin{equation}
\label{eq15a}
k_i = (3\pi /2)\rho (1+\sigma) c^2 r_c (r_c /h)^3 \, .
\end{equation}
Thus, if $r_c \approx h$, we obtain
$k_i \sim \rho c^2 \sqrt{A_i}$,
where $A_i = \pi r_c^2$ is the contact area.

%==================================================================
% \appendix
\section{Two simple examples of steady state solutions}
\label{simple}

\begin{figure}[h] %[t] \bigskip
\includegraphics[clip, width=8cm]{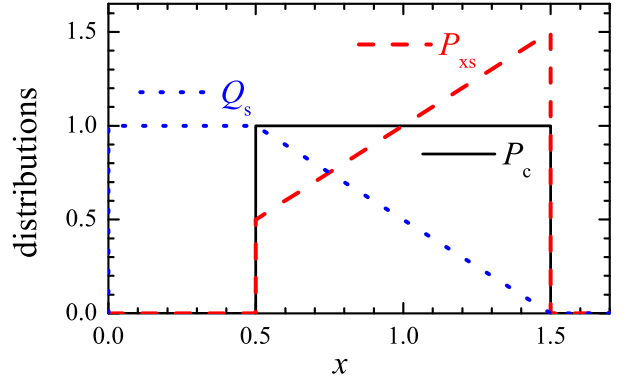}
\caption{\label{A02d}(Color online):
The stationary distribution $Q_s (x)$ (blue dotted line), and
the corresponding,
numerically obtained, distribution of static thresholds $P_{xs} (x)$
(red broken line)
for the rectangular distribution $P_c (x)$ (solid line).
%[use A02d]
}
\end{figure}
As a simple example, let us consider the distribution $P(x)=p \;
\Theta(x-x_s)$, or
\begin{equation}
P(x)=
\left\{
\begin{array}{c}
0 \;\;\; {\rm if} \;\;\; x \leq x_s,
\\
p \;\;\; {\rm if} \;\;\; x > x_s.
\end{array}
\right.
\label{Q14a}
\end{equation}
One can easily find that $C=x_s + p^{-1}$,
\begin{equation}
E_P (x)=
\left\{
\begin{array}{ccc}
1 & {\rm for} & 0 \leq x \leq x_s,
\\
e^{-p(x-x_s)} & {\rm for} & x > x_s,
\end{array}
\right.
\label{Q14b}
\end{equation}
\begin{equation}
P_c (x)=
\left\{
\begin{array}{ccc}
0 & {\rm for} & 0 \leq x \leq x_s,
\\
p \, e^{-p(x-x_s)} & {\rm for} & x > x_s,
\end{array}
\right.
\label{Q14d}
\end{equation}
and
\begin{equation}
F_k = {1 \over 2} N_c \langle k_i \rangle \left[ x_s + p^{-1} + p^{-1}
  /(1+p x_s) \right].
\label{Q14c}
\end{equation}
In particular, if $x_s =0$, then $Q_s (x)= \Theta(x) p \, e^{-px}$
and $F=N_c \langle k_i \rangle /p$,
while for the case of $x_s >0$ and $p \to \infty$ we obtain that
$Q_s (x)=x_s^{-1}$ within the interval $0 \leq x \leq x_s$ and 0 outside it,
so that $F = {1 \over 2} N_c \langle k_i \rangle x_s$.

%-------------------------------------------------------------------
\smallskip
Another simple example, which admits an exact solution,
is the case of the rectangular $P_c (x)$ distribution shown in Fig.~\ref{A02d},
\begin{equation}
P_c (x)= \left\{ \begin{array}{ll}
0 & {\rm if} \;\;\; x \leq 0.5,
\\
1 & {\rm if} \;\;\; 0.5 < x < 1.5,
\\
0 & {\rm if} \;\;\; x \geq 1.5.
\end{array}
\right.
\label{Q22a}
\end{equation}
The fraction density of breaking contacts $P(x)$ is given by Eq.~(\ref{PcP}):
\begin{equation}
P(x)= \left\{ \begin{array}{cl}
0 & {\rm if} \;\;\; x \leq 0.5,
\\
(1.5-x)^{-1} & {\rm if} \;\;\; 0.5 < x < 1.5,
\\
\infty & {\rm if} \;\;\; x \geq 1.5.
\end{array}
\right.
\label{Q22b}
\end{equation}
$P(x)$, given by Eq.~(\ref{PcP}) diverges when $x$ tends to $1.5$, and
then, according to its physical meaning it has to be infinite for
larger values of $x$.
In this case
\begin{equation}
U(x)= \left\{ \begin{array}{cl}
0 & {\rm if} \;\;\; x \leq 0.5,
\\
-\ln (1.5-x) & {\rm if} \;\;\; 0.5 < x < 1.5,
\\
\infty & {\rm if} \;\;\; x \geq 1.5,
\end{array}
\right.
\end{equation}
\begin{equation}
E_P (x)= \left\{ \begin{array}{cl}
1 & {\rm if} \;\;\; x \leq 0.5,
\\
1.5-x & {\rm if} \;\;\; 0.5 < x < 1.5,
\\
0 & {\rm if} \;\;\; x \geq 1.5.
\end{array}
\right.
\end{equation}

The earthquake model with the distribution (\ref{Q22a})
and the master equation model with $P(x)$ given by  Eq.~(\ref{Q22b})
have \textit{exactly} the same solution for any initial configuration.

Notice that this example clearly demonstrates
that the probability distribution $P_c (x)$,
which determines the static thresholds
for the newborn contacts,
is different from the distribution of
static thresholds $P_{xs} (x)$ which is achieved in the steady state as
pointed out in Sec.~\ref{earthquakes}.

%==================================================================
\section{Friction loop}
\label{loop}

\begin{figure}[h] %[t] \bigskip
\includegraphics[clip,width=8cm]{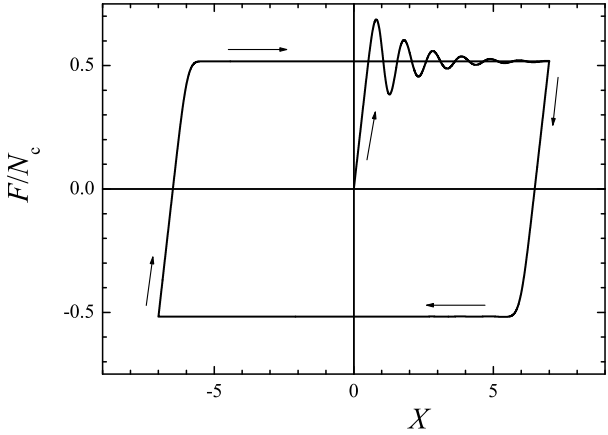} % clip
\caption{\label{A09}
Friction loop.
The distribution $P_c (x)$ is Gaussian with
$\bar{x}_s =1$ and $\sigma_s =0.2$,
the initial distribution $Q_{\rm ini} (x)$ is
Gaussian with $\bar{x}_{\rm ini} =0$ and $\sigma_{\rm ini} =0.025$.
The top substrate moves to the right, then to the left, and finally
again to the right
as indicated by arrows.
%[use Rotate/idl/earth14d.pro = A09]
}
\end{figure}

Through the paper we assumed that the top substrate moves continuously
to the right, i.e.\ $\Delta X >0$.
When the substrate moves to the left, or $\Delta X <0$,
Eqs.~(\ref{Q3}), (\ref{Q4}) and~(\ref{Q5a}) must be modified in the
following manner:
\begin{equation}
\Delta Q_-(x;X) = -P_b (x) \, \Delta \! X \, Q(x;X)  \; ,
\label{Q3back}
\end{equation}
\begin{equation}
\Delta Q_+(x;X) = R_b(x) \int_{-\infty}^{\infty}
dx^{\prime} \, \Delta Q_-(x^{\prime};X)
\label{Q4back}
\end{equation}
and
\begin{eqnarray}
\frac{ \partial Q(x;X) }{ \partial x} + \frac{ \partial Q(x;X) }{ \partial X} -
P_b(x) \, Q(x;X)
\nonumber \\
= R_b(x) \int_{-\infty}^{\infty} dx^{\prime} \, P_b (x^{\prime}) \, Q(x^{\prime};X),
\label{Q5aback}
\end{eqnarray}
where we have denoted by index $b$ ($P_b(x)$, $R_b(x)$) quantities relative
to the backward motion.
Comparing the equations for the forward and backward motions we see that
they obey the symmetry relations $P_b (x) = P(-x)$ and $R_b(x) =
R(-x)$ which are a manifestation of the irreversibility of the master equation.
Indeed, if the force $f_i$ on a given contact $i$
approaches and overcomes $f_{si}$,
the contact breaks; but if we now reverse the direction of the motion, this
contact does not jump back
to the value $f_{i} \approx f_{si}$; instead $f_i$ decreases until
it reaches the value $f_{i} \approx -f_{si}$.

Although this comment does not bring in new physics, it is important
for experimentalists.  In a typical tribological experiment, the top
substrate periodically moves forward and backward, and as a result, the
so-called ``friction loop'' is observed.  The same loop can easily be
calculated with the ME approach as shown in Fig.~\ref{A09}.

%==================================================================
\section{Delay effects}
\label{delay}

\begin{figure}[h] %[t] \bigskip
\includegraphics[clip,width=8cm]{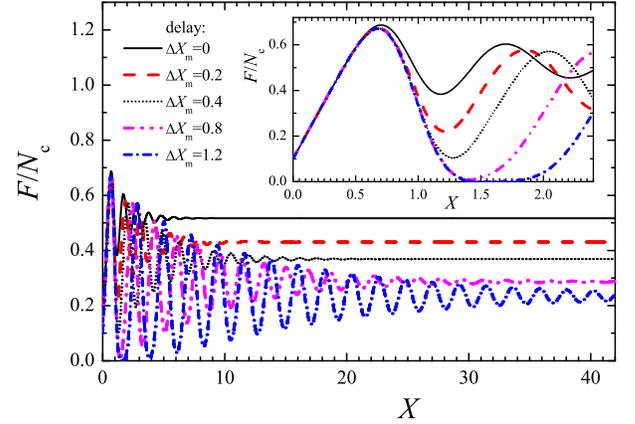} % clip
\caption{\label{A13a} (Color online):
The dependences $F(X)$ for different delay times
$\Delta X_m =0$ (black solid),
0.2 (red dashed),
0.4 (black dotted),
0.8 (magenta dash-dot-dotted) and
1.2 (blue dash-dotted curve).
The distribution $P_c (x)$ is Gaussian with
$\bar{x}_s =1$ and $\sigma_s =0.2$,
the initial distribution $Q_{\rm ini} (x)$ is
Gaussian with $\bar{x}_{\rm ini} =0.1$ and $\sigma_{\rm ini} =0.025$,
and $\langle k_i \rangle =1$.
%[use earth16.pro = A13a]
}
\end{figure}

\begin{figure}[t] %[t] \bigskip
\includegraphics[clip,width=8cm]{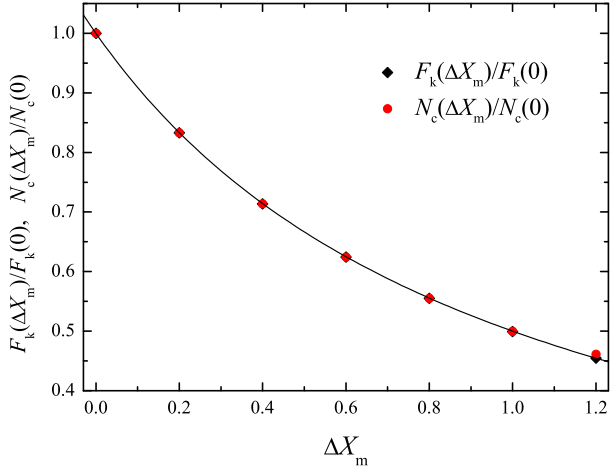} % clip
\caption{\label{A13b} (Color online):
The final number of asperities in contact (red circles)
and the kinetic friction force (black diamonds)
in smooth sliding regime as functions of the delay time
(other parameters as in Fig.~\ref{A13a}).
The solid curve shows the $(1+ \Delta X_m)^{-1}$ dependence.
%[use earth16.pro = A13b]
}
\end{figure}

To establish Eq.~(\ref{Q4}) we assumed that, after a slip, a contact
would stick again without any delay. Actually restoring the contact
always requires a small delay $\tau_m$, particularly for the melting/freezing
mechanism.
Simulation~\cite{BN2006} shows that $\tau_m \sim 10^2 \tau_0$, where
$\tau_0 \sim 10^{-13}$s is a characteristic period of atomic
oscillations in the lubricant. Therefore to
write Eq.~(\ref{Q4}) we have to use $\Delta Q_+(x;X + \Delta \! X_m)$,
where $\Delta \! X_m = v \tau_m$ and $v$ is the driving velocity.

In the body of paper we ignored the delay effect.
In this case $\Delta \! X_m = \Delta \! X$,
and we may drop the $+ \Delta \! X$ because it leads to a second order
correction since it appears in a term which is already a correction to $Q$.

When the delay is taken into account,
the integro-differential equation (\ref{Q5a}) has to be modified to become
\begin{eqnarray}
\frac{ \partial \widetilde{Q}(x;X) }{ \partial x} +
\frac{ \partial \widetilde{Q}(x;X) }{ \partial X} + P(x) \, \widetilde{Q}(x;X)
\nonumber \\
= R(x) \int_{-\infty}^{\infty} dx^{\prime} \, P(x^{\prime}) \,
\widetilde{Q}(x^{\prime};X - \Delta \! X_m).
\label{Q5adelay}
\end{eqnarray}
The normalization condition (\ref{Q5c}) does not hold for
$\widetilde{Q}(x;X)$ because the total number of  unbroken contacts,
is not conserved anymore.
When an asperity breaks it is not in contact with the substrate
during the time $\tau_m$  until the contact is restored.
Thus, $N_c$ varies with $X$, and $N_c (X) < N_c (0)$,
because some contacts are in a temporarily broken state even at smooth sliding.
Typical dependences $F(X)$ for different values of the delay time are
shown in Fig.~\ref{A13a}.
The kinetic friction force during smooth sliding is such that
$F_k \propto N_c (\infty)$ and, therefore, it depends on pre--history of
the contacts.
If one starts from the same initial configuration,
the final force $F_k$ is lower when the delay
time increases (see Fig.~\ref{A13b}).
The dependence $F_k (v)$ is described by Eq.~(\ref{fitFv})
with $F(\infty) =0$ and $v_{\rm aging} = \bar{x}_s /\tau_m$.

%==================================================================
\section{The singular case}
\label{singular}

\begin{figure*}[t] %[ht] \bigskip
\includegraphics[width=8.5cm, height=4cm]{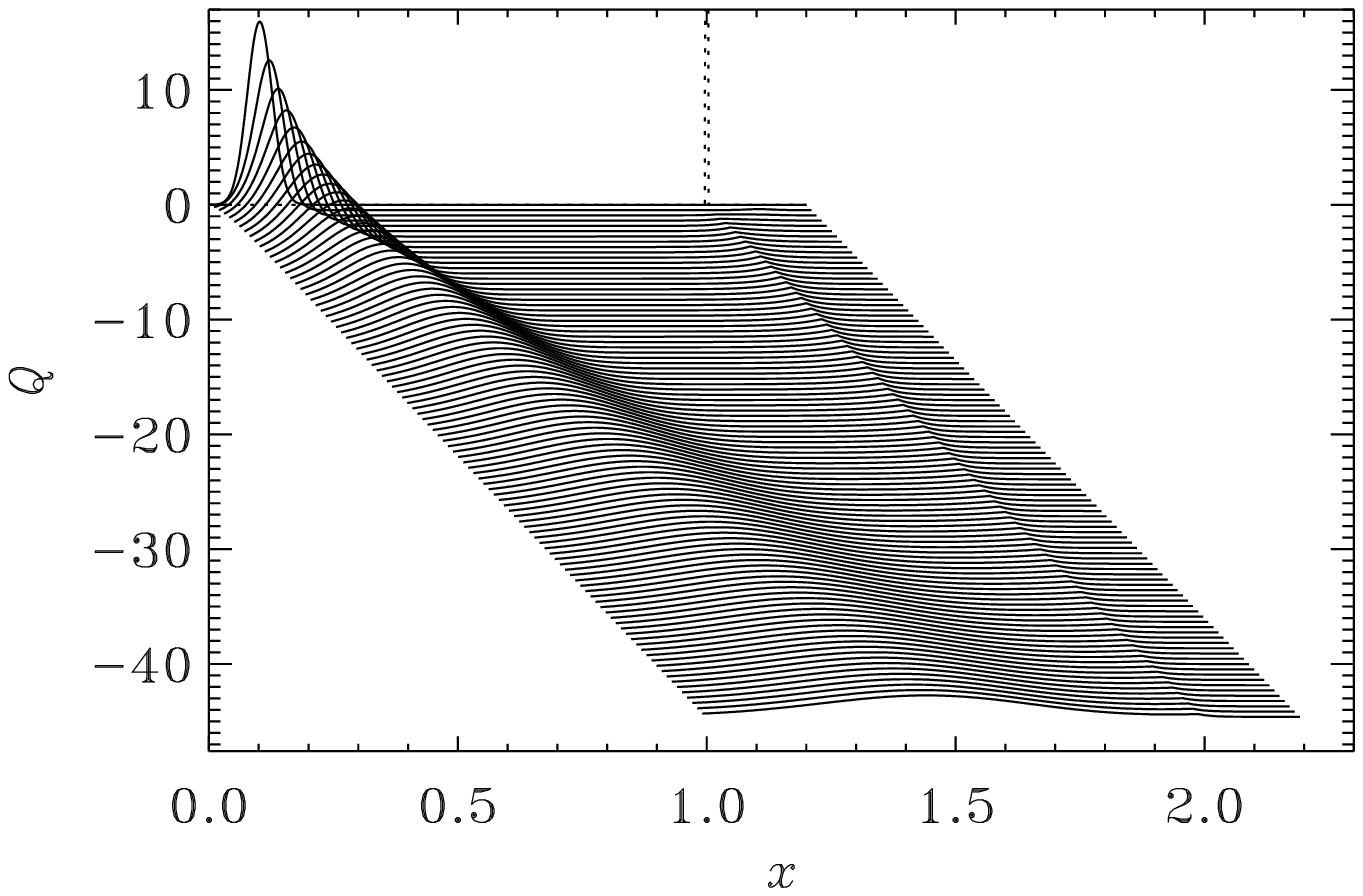} % clip
\hspace{0.5cm}
\includegraphics[width=8.5cm, height=4cm]{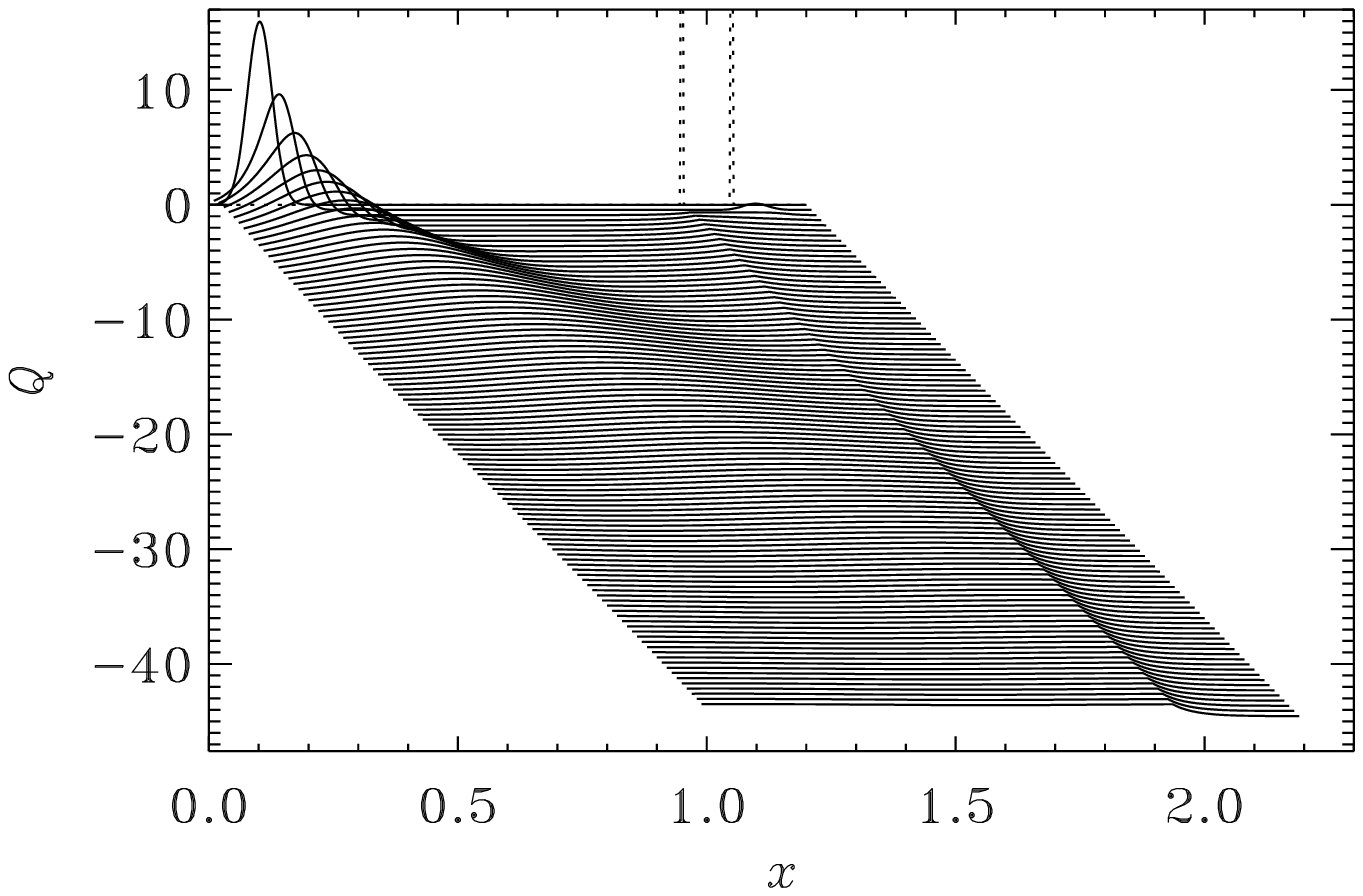} % clip
\includegraphics[width=8.4cm]{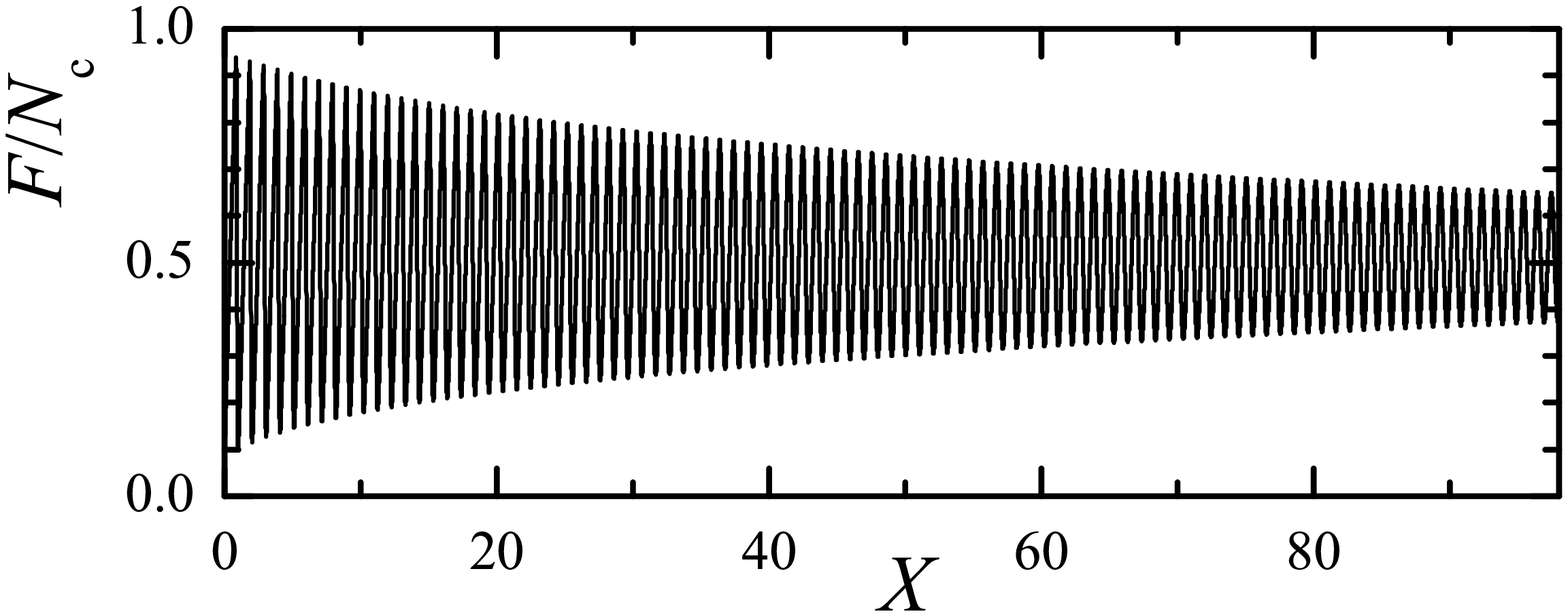}
\hspace{0.7cm}
\includegraphics[width=8.4cm]{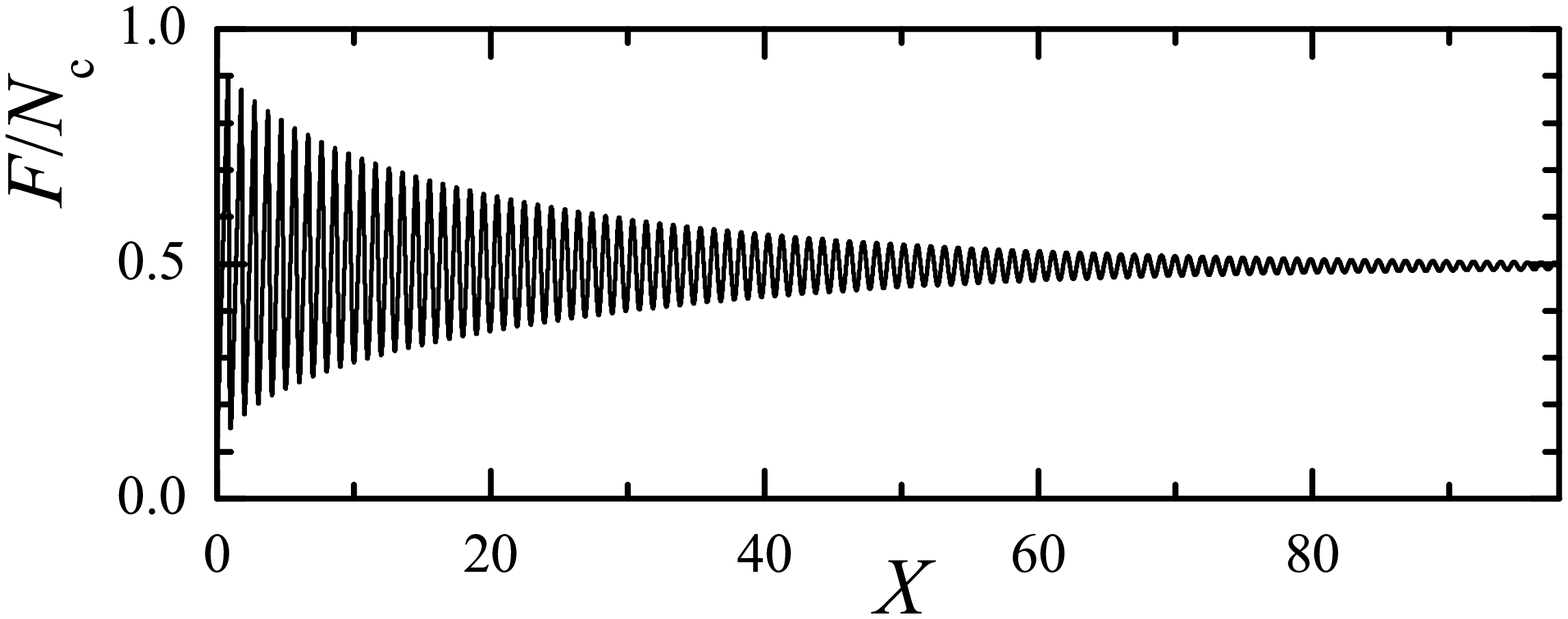}
\caption{\label{delta12}
Evolution of the model for two $P_c (x)$ distributions:
Left column is for a one-peak distribution of threshold,
$P_c (x) = G (x; \bar{x},\sigma))$ with $\bar{x} =1$ and $\sigma =10^{-3}$
(i.e., close to the delta-function distribution),
while right column is for a two-peak distribution of threshold,
$P_c (x) = {1 \over 2} \left[ G (x; 0.95 \bar{x},\sigma) + G (x; 1.05
  \bar{x},\sigma) \right]$
(i.e., close to the distribution with two delta-functions).
Solid curves show the distribution $Q(x;X)$ for incrementally
increasing values of $X$
with the increment $\Delta X \approx 1$,
dotted curves show $P_c(x)$ (for $X=0$ only).
The bottom row shows the corresponding dependences $F(X)$ for $\langle
k_i \rangle =1$ and $K=\infty$.
The initial distribution $Q_{\rm ini} (x)$ is
Gaussian with $\bar{x}_{\rm ini} =0.1$ and $\sigma_{\rm ini} =0.025$.
%[use Rotate/idl/delta]
}
\end{figure*}
As we have shown, in a general case the solution of the master
equation always approaches the steady-state solution given by
Eqs.~(\ref{Q11a}--\ref{Q11c}),
which describes the smooth sliding.
However, there is one exception from this general scenario,
when the model admits a periodic solution and $F(X) \neq$~Const even
in the limit $X \to \infty$.
This is the {\it singular\/} case
when all contacts are identical, i.e., all contacts are characterized by
the same static threshold $x_s$,
so that $P_c(x) = \delta (x-x_s)$, or $P(x)=0$ for $x<x_s$ and
$P(x)=\infty$ for $x \geq x_s$.

As one can check by direct substitution, in this particular case
the steady-state solution of Eq.~(\ref{Q5a}) has the following form:
\begin{equation}
Q(x;X) = S(x-X)
\left[ \Theta(x) - \Theta(x-x_s) \right],
\label{Q7b}
\end{equation}
where the function $S(\xi)$ is defined by the initial condition:
$S(\xi) = Q_{\rm ini} (\xi)$ for the interval $0 \leq \xi < x_s$,
and then $S(\xi)$ should be repeated periodically over the whole interval
$-\infty \leq \xi < \infty$,
\[ S(\xi \pm n x_s) = S(\xi) \quad \forall \; n~{\mathrm{integer}} \; . \]
In this simple case the total force (\ref{Q6}) is equal to
\begin{equation}
F(X) = N_c k \left[ X + \int_{-X}^{x_s -X} d\xi \, \xi S(\xi) \right].
\label{Q6e}
\end{equation}
The static friction force takes the minimal value $F_s = {1 \over 2} N_c k x_s$
for the uniform initial distribution, $Q_{\rm ini} (x) = x_s^{-1}$,
when $F(X)$ does not depend on $X$,
and the maximal value $F_s = N_c k x_s$ for the delta-function initial
distribution
$Q_{\rm ini} (x) = \delta (x - x_0)$ with some $0 \leq x_0 < x_s$,
when the function $F(X)$ has sawtooth shape changing from~0 to~$F_s$.

\smallskip
However, the periodic solution described above \textit{only} exists
for a distribution $P_c (x)$ with a single threshold.
If the contacts are characterized by more than one threshold value, for example,
if one part of contacts has the threshold $x_{s1}$ and
another part the threshold $x_{s2} \neq x_{s1}$
[i.e., $P_c (x)$ is described by a sum of two delta-functions],
then the system will always approach the stationary steady state.
This is demonstrated in Fig.~\ref{delta12}, where we compare the
system evolution in cases of one-peak and two-peak $P_c (x)$ distributions.
Notice, however, that this statement is only valid for an infinite
set of contacts
(the number of contacts with each threshold must be infinite)
and cannot be applied for a microscopic system where, e.g.
two tips move over a surface.

%==================================================================


\begin{thebibliography}{xx}

\bibitem{P0}
  B.N.J.\ Persson, {\it Sliding Friction: Physical Principles and
    Applications} (Springer-Verlag, Berlin, 1998).
\bibitem{MUR2003}
  M.H.\ M\"user, M.\ Urbakh, and M.O.\ Robbins, Adv.\ Chem.\ Phys.\ {\bf 126}, 187 (2003).
\bibitem{BN2006}
  O.M.\ Braun and A.G.\ Naumovets, Surf.\ Sci.\ Reports {\bf 60}, 79 (2006).
\bibitem{BC2006}
  T.\ Baumberger and C.\ Caroli, Advances in Physics {\bf 55}, 279 (2006).
\bibitem{RT1991}
  M.O.\ Robbins and P.A.\ Thompson, Science {\bf 253}, 916 (1991).
\bibitem{BP2001}
  O.M.\ Braun and M.\ Peyrard, \pre {\bf 63}, 046110 (2001).
\bibitem{BR2002}
  O.M.\ Braun and J.\ R\"oder, \prl {\bf 88}, 096102 (2002).
\bibitem{BPBFV2005}
  O.M.\ Braun, M.\ Peyrard, V.\ Bortolani, A.\ Franchini, and
  A.\ Vanossi, \pre {\bf 72}, 056116 (2005).
\bibitem{pheno}
  F.\ Heslot, T.\ Baumberger, B.\ Perrin, B.\ Caroli, and C.\ Caroli,
  \pre {\bf 49}, 4973 (1994);
  T.\ Baumberger, C.\ Caroli, B.\ Perrin, and O.\ Ronsin, \pre {\bf
    51}, 4005 (1995).
\bibitem{P1997}
  B.N.J.\ Persson, \prb {\bf 55}, 8004 (1997).
\bibitem{BK1967}
  R.\ Burridge and L.\ Knopoff, Bull.\ Seismol.\ Soc.\ Am.\ {\bf 57},
  341 (1967).
\bibitem{CL1989}
  J.M.\ Carlson and J.M.\ Langer, \prl {\bf 62}, 2632 (1989).
\bibitem{OFC1992}
  Z.\ Olami, H.J.S.\ Feder, and K.\ Christensen, \prl {\bf 68}, 1244 (1992).
\bibitem{P1995}
  B.N.J.\ Persson, \prb {\bf 51}, 13568 (1995).
\bibitem{FKU2004}
  A.E.\ Filippov, J.\ Klafter, and M.\ Urbakh, \prl {\bf 92}, 135503
  (2004).
\bibitem{FDW2005}
  Z.\ Farkas, S.R.\ Dahmen, and D.E.\ Wolf, J.\ Stat.\ Mech.: Theory
  and Experiment P06015 (2005);
  cond-mat/0502644.
\bibitem{BP2008} O.M.\ Braun and M.\ Peyrard, \prl {\bf 100}, 125501
  (2008).
\bibitem{BT2009}
  O.M.\ Braun and E.\ Tosatti, Europhys. Lett.\ {\bf 88}, 48003 (2009).
\bibitem{BBU2009}
  O.M.\ Braun, I.\ Barel, and M.\ Urbakh, \prl {\bf 103}, 194301 (2009).
\bibitem{YCI1993}
  H.\ Yoshizawa, Y.\ L.\ Chen, and J.\ Israelachvili, Wear {\bf 168},
  161 (1993);
  H.\ Yoshizawa and J.\ Israelachvili, J.\ Phys.\ Chem.\ {\bf 97},
  11300 (1993).
\bibitem{GW1966}
  J.A.\ Greenwood and J.B.P.\ Williamson, Proc.\ Roy.\ Soc.\ A {\bf
    295}, 300 (1966).
\bibitem{Garg1995}
  A.\ Garg, \prb {\bf 51}, 15592 (1995).
\bibitem{Dudko2003}
  O.K.\ Dudko, A.E.\ Filipov, J.\ Klafter, and M.\ Urbakh, PNAS {\bf
  100}, 11378 (2003).
\bibitem{SW2009}
  M.\ Srinivasan and S.\ Walcott, \pre {\bf 80}, 046124 (2009).
\bibitem{YGI1993}
  H.\ Yoshizawa, P.\ McGuiggan, and J.\ Israelachvili, Science {\bf
    259}, 1305 (1993).
\bibitem{SO2003}
  S.\ Sills and R.M.\ Overney, \prl {\bf 91}, 095501 (2003).
\bibitem{SJHF2006}
  A.\ Schirmeisen, L.\ Jansen, H.\ H\"olscher, and H.\ Fuchs,
  Apl.\ Phys.\ Lett.\ {\bf 88}, 123108 (2006).
\bibitem{MIKOT2005}
  A.\ Maeda, Y.\ Inoue, H.\ Kitano, S.\ Okayasu, and I.\ Tsukada,
  Int.\ J.\ Mod.\ Phys.\ B {\bf 19}, 463 (2005).
\bibitem{MIKSOTN2005}
  A.\ Maeda, Y.\ Inoue, H.\ Kitano, S.\ Savel'ev, S.\ Okayasu,
  I.\ Tsukada, and F.\ Nori,
  \prl {\bf 94}, 077001 (2005).
\bibitem{MN2007}
  A.\ Maeda and D.\ Nakamura,
  Journal of Physics: Conference Series {\bf 89}, 012020 (2007).
\bibitem{NKKGMKB2007}
  D.\ Nakamura, T.\ Kubo, S.\ Kitamura, L.B.\ G\'omez, A.\ Maeda,
  M.\ Konczykowski, and C.J.\ van der Beek,
  Journal of Physics: Conference Series {\bf 89}, 012021 (2007).
\bibitem{J1985}
  K.L.\ Johnson, \textit{Contact Mechanics} (Cambridge University
  Press, Cambridge, 1985).
\bibitem{P2001a}
  B.N.J.\ Persson, J.\ Chem.\ Phys.\ {\bf 115}, 3840 (2001).
\bibitem{YP2008}
  C.\ Yang and B.N.J.\ Persson, J.\ Phys.: Condens.\ Matter {\bf 20},
  215214 (2008).
\bibitem{HS1993}
  M.\ Hirano and K.\ Shinjo, Wear {\bf 168}, 121 (1993).
\bibitem{RS1996}
  M.O.\ Robbins and E.D.\ Smith, Langmuir {\bf 12}, 4543 (1996).
\bibitem{QCCG2002}
  Y.\ Qi, Y.-T.\ Cheng, T.\ Cagin, and W.A.\ Goddard III, \prb {\bf
    66}, 085420 (2002).
\bibitem{DVPFHZ2004}
  M.\ Dienwiebel, G.S.\ Verhoeven, N.\ Pradeep, J.W.M.\ Frenken,
  J.A.\ Heimberg, and H.W.\ Zandbergen, \prl {\bf 92}, 126101 (2004).
% \bibitem{B1990}
%   O.M.\ Braun, Surface Sci.\ {\bf 230}, 262 (1990).
\bibitem{BM2010b}
  N.\ Manini and O.M.\ Braun, \pre, submitted (manuscript ET10772) (2010).
\bibitem{JHT2006}
  A.\ Jabbarzadeh, P.\ Harrowell, and R.I.\ Tanner, \prl {\bf 96},
  206102 (2006).
\bibitem{JHT2005}
  A.\ Jabbarzadeh, P.\ Harrowell, and R.I.\ Tanner, \prl {\bf 94},
  126103 (2005).
\bibitem{HR2001s}
  G.\ He and M.O.\ Robbins,   \prb {\bf 64}, 35413 (2001).
\bibitem{HR2001k}
  G.\ He and M.O.\ Robbins, Tribology Letters {\bf 10}, 7 (2001).
\bibitem{BM2010a}
  O.M.\ Braun and N.\ Manini, \pre, submitted (manuscript EM10550) (2010).
\bibitem{LS1958}
  E.M.\ Lifshitz and L.P.\ Pitaevskii, \textit{Physical Kinetics}
  (Pergamon, Oxford, 1981).
\bibitem{HS1971}
  A.F.\ Huxley and R.M.\ Simmons, Nature (London) {\bf 233}, 533 (1971).
\bibitem{P1975a}
  C.S.\ Peskin, \textit{Lectures on Mathematical Aspects of Physiology}
  (Courant Institute of Mathematical Sciences, New York, 1975).
\bibitem{P1975b}
  C.S.\ Peskin, \textit{Mathematical Aspects of Heart Physiology}
  (Courant Institute of Mathematical Sciences, New York, 1975).
\bibitem{LP1986}
  H.M.\ Lacker and C.S.\ Peskin, Lect.\ Math Life Sci.\ (American
  Mathematical Society,
  Providence, RI, 1986), Vol. 16, p. 121.
\bibitem{W1984}
  W.O.\ Williams, Mathematical Biosciences {\bf 70}, 203 (1984).
\bibitem{LL1986}
  L.D.\ Landau and E.M.\ Lifshitz, \textit{Theory of Elasticity},
  Course of Theoretical Physics Vol.\ 7 (Pergamon, New York, 1986).

\end{thebibliography}
\end{document}